\documentclass[10pt,twocolumn,twoside]{IEEEtran}  
\IEEEoverridecommandlockouts     
\usepackage{siunitx}

\usepackage{algorithm,algorithmic}
\usepackage{graphicx}
\usepackage{textcomp}
\usepackage{xcolor}
\usepackage{graphics} 
\usepackage[outdir=./]{epstopdf}
\usepackage{subcaption}
\usepackage{stfloats}

\usepackage{pgfplots}
\pgfplotsset{compat=newest}
 \usetikzlibrary{plotmarks}
 \usetikzlibrary{arrows.meta}
 \usepgfplotslibrary{patchplots}
 \usepackage{grffile}
\usepackage{tikz}
\usetikzlibrary{shapes, fit, decorations.markings, matrix, plotmarks, positioning, fit, spy, patterns, shadows, calc, backgrounds, arrows.meta,spy}
\usepackage[american]{circuitikz} 

\tikzstyle{place}=[circle,draw=black!50,fill=mustard, thick,  inner sep=0pt,minimum size=8mm]
\tikzstyle{connection}=[circle,fill=black,inner sep=1.5pt]

\definecolor{node_colour}{HTML}{009999}
\definecolor{mycolor1}{rgb}{0.00000,0.44700,0.74100}%
\definecolor{mycolor2}{rgb}{0.85000,0.32500,0.09800}%
\definecolor{mycolor3}{rgb}{0.92900,0.69400,0.12500}%
\definecolor{mycolor4}{rgb}{0.49400,0.18400,0.55600}%
\definecolor{mycolor5}{rgb}{0.46600,0.67400,0.18800}%
\definecolor{mycolor6}{rgb}{0.30100,0.74500,0.93300}%
\definecolor{mustard}{RGB}{7,194,201} 

\tikzstyle{place}=[circle,draw=black!50,fill=node_colour, thick,  inner sep=0pt,minimum size=8mm]
\tikzstyle{connection}=[circle,fill=black,inner sep=1.5pt]
\tikzstyle{vertex1}=[scale=1,draw,circle,fill=node_colour]
\tikzstyle{edge} = [draw,-]

\usepgfplotslibrary{fillbetween}

\usepackage{mathtools,amsmath,amssymb,amsfonts,mathrsfs,amsthm,mathdots}
\usepackage{commath}
\usepackage{bbm}
\usepackage{dsfont}
\usepackage{nicematrix}

\newtheorem{theorem}{Theorem}
\newtheorem{corollary}{Corollary}
\newtheorem{lemma}{Lemma}
\newtheorem{proposition}{Proposition}
\newtheorem{assumption}{Assumption}

\theoremstyle{definition} 
\newtheorem{definition}{Definition}
 
\theoremstyle{remark}
\newtheorem{remark}{Remark}

\DeclareCaptionLabelFormat{algnonumber}{Algorithm}
\newcommand{\ts}[1]{{\textnormal{#1}}}

\newcommand{\ie}{\emph{i.e.}\ }

\newcommand{\eg}{\emph{e.g.}\ }

\newcommand{\mc}{\mathcal}

\newcommand{\mbb}{\mathbb}
\newcommand{\mf}{\mathfrak}

\newcommand\st{\colon~}

\newcommand{\innerproduct}[2]{\langle #1, #2 \rangle}

\usepackage{cite}
\title{Decentralized Voltage Control of AC Microgrids with Constant Power Loads using Control Barrier Functions}
\author{Grigoris Michos, and George C. Konstantopoulos
\thanks{This work was performed within the TARGET-X framework, a project funded by the Smart Networks and Services Joint Undertaking (SNS JU) under Horizon Europe (funding number 101096614). The authors are with the
Department of Electrical and Computer Engineering, University of Patras, Rion 26504, Greece.
        {\tt\small (e-mail: grmichos@upatras.gr, \\ g.konstantopoulos@ece.upatras.gr)}}%
}

\begin{document}
\maketitle

\begin{abstract}
This paper proposes a novel nonlinear decentralized voltage controller for constrained regulation of meshed AC Microgrid networks with high penetration of time-varying constant power loads.  Modelling the load demand as a constantly evolving unknown disturbance, the network model is reformulated in a cascaded structure composed of a nominal, \ie uncertainty-free, and an error subsystem.  By adopting a suitable control barrier function, we formulate a continuous-time control law and derive analytic conditions on the tuning parameters, such that the distance between the true and the nominal state trajectories is bounded. Under sufficient conditions, we prove asymptotic stability of the cascaded dynamics with respect to an equilibrium set and also provide an estimate of the region of attraction. In addition, it is rigorously shown that the proposed nonlinear control law enforces constrained regulation around a rated voltage value, without the need of saturation devices. The operation of the closed-loop system is illustrated both via simulation  and real-time HIL scenarios, demonstrating bounded operation and convergence to a neighbourhood of the desired reference vector.
\end{abstract}


\section{Literature Review}

The modern power grid is undergoing a transformative evolution from a traditionally centralized infrastructure to a decentralized entity. The concept of Microgrid has emerged, in order to facilitate the decentralized regulation of various distributed energy resource (DER) units and enable their integration with the main power grid. However, lack of synchronous generators (SGs) and rotational kinetic energy has a negative effect on the total system inertia, increasing susceptibility to surges in power demand and faults \cite{dorfler2023control}.

A Microgrid has the ability to operate both in grid-connected and islanded modes. In the former, local converters operate in a grid-following mode, where the voltage levels are dictated by a strong grid. In the latter, both the voltage and the frequency are regulated by the so-called grid-forming converters. The second case is particularly challenging due to the lack of SGs and their ability to store large amounts of energy within their rotational movement. Conventionally, voltage control for islanded Microgrids is performed with the droop control, where the inner loops regulation is achieved by linear PI controllers \cite{chandorkar2002control}. Other control strategies, \eg the Virtual Synchronous Machine control, involve emulation of the SG \cite{d2015virtual} that can provide reliable grid support but require careful tuning of the inner loops to avoid instability. In an effort to achieve stronger stability guarantees, the dispatchable virtual oscillator was proposed, see for example \cite{gross2019effect}, where it is shown that for some network topologies it is possible to achieve (semi) global asymptotic stability certificates.

Two main categories exist in the literature, that investigate the stability of grid-forming converters, and by extension islanded Microgrid configurations. The majority of the proposed studies fall within the first, namely studies that numerically validate the stability certificates of the proposed closed-loop system \cite{alfaro2021distributed,khan2022robust}. The second category involves analytic approaches that either study the system behaviour locally in a neighbourhood of an operating point \cite{mohiuddin2019droop,markovic2021understanding} or attempt to achieve global results in a large-signal sense \cite{he2023nonlinear,subotic2020lyapunov}. However, to the best of the authors' knowledge, a limited number of studies investigate constant power demand in the network. Contrary to the literature of DC Microgrids, where the concept of constant power loads (CPLs) has being gaining increasing attention over the past few years, see for example \cite{michos2023robust,michos2024dynamic,braitor2024voltage}, in the case of AC Microgrid systems there is a large gap of control-theoretic studies considering the connection of tightly controlled power converters or rectifiers to the AC network.  At the same time, there exists an increasingly large number of electronic devices that require interconnection to Microgrids via power inverters  e.g. EV charger facilities, data centres or electronic motors that demand a constant power supply \cite{areerak2017adaptive,al2017constant}. In such cases, the AC side is required to meet a constant power demand, which, similarly to the case of a DC Microgrid, can significantly degrade the system stability. Numerical or local approximation approaches have been proposed in \cite{molinas2008constant,liu2012novel,molinas2008investigation}, however these require the knowledge of the network parameters and do not consider the nonlinear characteristics of the system. The system dynamic model was considered in \cite{areerak2017adaptive} but the stability certificates are only numerically investigated. An infinity-norm criterion was established in \cite{liu2014infinity}, but is limited in providing results for a single inverter feeding a constant power load, while the analysis also omits the system nonlinearities. Overall, the topic of AC Microgrids with large penetration of constant power loads is largely unexplored and, to the best of our knowledge, no analytic results have been established that provide system stability in a rigorous fashion.

A second important topic is the strict voltage limitation within a desired operating region. Considering the low system inertia and the lack of sufficient energy storage capacity of inverter-based networks, the system becomes vulnerable to power surges and voltage drops. Generally, the task of constraint satisfaction is achieved by a supervisory controller, which only addresses steady-state performance \cite{mansoorhoseini2022islanded,zhou2021distributed}. A few studies have proposed controller schemes that incorporate this requirement in the primary control level; the authors of \cite{KOLSCH202012229} propose a model-based voltage controller for AC Microgrids that tracks reference points provided by solving a constrained optimization problem. In \cite{mottaghizadeh2022robust}, a model-predictive controller established constrained voltage regulation and robustness with respect to system disturbances. However, the former study is limited to steady-state bounds, while the latter requires knowledge of the network parameters. In addition, both approaches impose a large computational load, especially in the case of AC Microgrids that operate in fast time scales. In \cite{11037272}, a nonlinear controller enforces an upper bound on the voltage trajectory, however the study only considers purely passive networks.  Finally, the conventional method to enforce boundedness of power systems' state trajectories involves the adoption of saturation devices on the reference values \cite{rocabert2012control,qoria2018tuning}, which however induce non-smooth closed loop dynamics and have been known to cause instabilities \cite{xin2016synchronous}. 

The concept of control barrier functions (CBFs) has risen in the literature of control, in an effort to enforce the system states to operate within a ``safe" region \cite{ames2016control,xu2015robustness}. The main benefit of this approach is that it is possible to establish set invariance under the system trajectories without requiring invariance of each sublevel set, \ie a necessary property of Lyapunov-based approaches. Recently, CBFs have begun gaining interest in the literature of power systems. A CBF-based supervisory controller was proposed in \cite{wu2024fixed} for inverter-based networks. The authors of \cite{hassan2024resilience} propose a CBF approach to ensure network resilience and voltage restoration to a ``safe" region. The regulation of inverter-based Microgrids is studied in \cite{kundu2019distributed}, where the proposed controller utilizes CBFs to enforce voltage constraints both during transient and steady state performance. While a promising approach, the application of CBFs on AC microgrids is largely unexplored, with many studies focusing on the DC counterpart. Nevertheless, CBFs can play a crucial role in designing stabilizing controllers for AC Microgrid networks that are required to satisfy CPL demand, where the dynamics are no longer globally Lipschitz continuous.

This the the the article aims to fill the gap identified in the literature, by proposing a local nonlinear voltage control law that stabilizes the network under time-varying CPL demand, and enforces constrained operation around a rated voltage value. Each controller is implemented locally at each interfacing inverter unit and regulates the local output capacitor AC voltage. The proposed study perceives the CPL demand as a system disturbance and proposes a tube-control approach, where the original dynamics are expressed by two states; a nominal, \ie disturbance-free, and an error between the true state and nominal. Then, the contributions of this work are 
\begin{itemize}
\item \textit{C1:} We propose a smooth, decentralized, continuous-time feedback control law that guarantees boundedness of the error dynamics within a predefined tube. Specifically, we consider the tube cross-section at any point in time as a set defined by an appropriate CBF. Then, we use set invariance theory to derive analytic conditions on the tuning parameters that guarantee a positive invariance of the tube cross-section, under the solution of the error dynamics, despite the presence of fluctuating load demand. Contrary to \cite{mansoorhoseini2022islanded,zhou2021distributed,habibi2021unfalsified,ullah2022voltage,KOLSCH202012229,mottaghizadeh2022robust} the proposed control design guarantees constraint satisfaction in a continuous-time fashion, without the need of optimization-based techniques that require both knowledge of the network parameters and impose a large computational load.
\item \textit{C2:} A novel, decentralized, bounded integral controller is proposed to regulate the nominal state to a desired a reference point. First, we incorporate the requirement for bounded RMS output voltage within the design procedure of the proposed integral controller. Then, we analytically derive conditions for boundedness of the nominal state trajectories, such that the output nodal RMS voltage is contained in a desired operating region.
\item \textit{C3:} We prove asymptotic stability of both the nominal and the overall network dynamics using Lyapunov theory and provide an estimate for the region of attraction (RoA) for the closed-loop system equilibrium set. To the authors' best knowledge, and contrary to \cite{areerak2017adaptive,molinas2008constant,liu2012novel,molinas2008investigation,liu2014infinity}, this is the first work that Lyapunov guarantees are derived by considering AC Microgrid networks with high penetration of CPLs. In addition, contrary to \cite{11037272}, we prove that the proposed approach achieves the aforementioned objectives, despite the presence of non-passive elements in the load model. 
\end{itemize}

The article is organized as follows. Sections \ref{sec:Prelims} and \ref{sec:Problem_Form} derive the system model and specify the control objectives. Section \ref{sec:Error} provides the set invariance analysis of the error dynamics. Boundedness and asymptotic stability of the nominal subsystem is proved in Section \ref{sec:Nominal}, while Section \ref{sec:Stability} establishes stability certificates for the cascaded network system. Section \ref{sec:Sims} includes illustrative examples of the closed-loop system, both via simulation and Hardware-In-Loop. Finally, Section \ref{sec:Conclusions} provides concluding remarks and future research directions.

\section{Preliminaries}\label{sec:Prelims}

\subsection{Notation}

For a vector $a \in \mbb{R}^n$, the $s^{\ts{th}}$ self-Hadamard product is denoted by $a^s$, while $\norm{a}$ denotes the euclidean norm unless otherwise specified. The notation $[a]$ denotes a diagonal matrix with diagonal elements $[a]_{ii} = a_i$. The $n \times n$ zero and identity matrices are denoted as $\mathbf{\ts{0}}_n$ and $\mathbf{\ts{I}}_n$ respectively. The Minkowski sum $A \oplus B$ of two sets $A \in \mbb{R}^n$ and $B \in \mbb{R}^n$ is defined as $A \oplus B \coloneqq \left\{ a+b \st a \in A, \ b \in B \right\}$. Consider the nonlinear system, $f\st \mbb{R}^n \to \mbb{R}^n$ given by
\begin{equation}\label{eq:bony}
\dot{x} = f(x).
\end{equation}
\begin{theorem}[Bony-Brezis \cite{redheffer1972theorems}]\label{thm:Bony}
Consider the nonlinear dynamics in (\ref{eq:bony}) and a closed subset $\mc{S} \subset \mbb{R}^n$. Let
\begin{enumerate}
\item $\left| f(x) - f(y) \right| \leq K|x-y|$, for some $K \in \mbb{R}$ and all $x,y \in \mc{S}$.
\item $\innerproduct{f(x)}{\nu(x)} \leq 0$, when $\nu(x) \in \mbb{R}^n$ is normal vector on $\mc{S}$ at $x$.
\end{enumerate}
Then, the set $\mc{S}$ is positive invariant under the solution of (\ref{eq:bony}).
\end{theorem}
\begin{definition}
Consider (\ref{eq:bony}), a continuously differentiable function $b \st \mbb{R}^n \to \mbb{R}$, and a set $\mc{S}$ defined by
\begin{subequations}
\begin{align}
\mc{S} \coloneqq \left\{x \in \mbb{R}^n \st b(x) \geq 0   \right\} \\
\partial \mc{S} \coloneqq \left\{x \in \mbb{R}^n \st b(x) = 0   \right\}
\end{align}
\end{subequations}
The function $b(\cdot)$ is called a control barrier function, if
\begin{equation}
\inf_{x \in \partial \mc{S}} \left\{ \mf{L}_f b(x) \right \} \geq 0,
\end{equation}
where $\mf{L}_f$ is the Lie derivative defined  by $\mf{L}_f \coloneqq \frac{\partial b(x)}{\partial x}f(x)$.
\end{definition}

\subsection{Network Modelling}
In this paper we investigate the problem of a Microgrid network consisting of $n$ DER units. Each $i$th unit, with $i \in \mc{M} \coloneqq \{1,2,3,\dots,n\}$, is interfaced with the network via 3-phase power inverters, local CPLs and primarily inductive lines. It is assumed that the injected current dynamics are operating in a sufficiently higher time-scale compared to the capacitor voltage, such that the injected current can be considered constant in the mathematical analysis of the capacitor voltage system. This is a common assumption in the literature, dating back to \cite{espinoza1992voltage}. In essence this allows us to consider the inverters as a controllable current source connected in parallel to an output capacitor. Exploiting the passivity of the lines, we can consider the network as a singularly perturbed system operating in two time scales \cite{Khalil2002a}. Then, similarly to \cite{baldivieso2020constrained}, adopting the Kirchhoff laws allows the derivation of the nodal d-q model in a synchronous rotating frame, rotating in a global frequency $\omega_g$, yielding 
\begin{subequations}\label{eq:True_System_Model_static}
\begin{align}
C_i \dot{v}_{d,i} &=I_{\ts{inj},di} + \omega_g C_i v_{q,i} - \mc{L}_i^{\top} v_{d} - g_{d,i}(v_{d,i},v_{q,i},P_i,Q_i), \\ 
C_i \dot{v}_{q,i} &= I_{\ts{inj},qi} - \omega_g C_i v_{d,i} -  \mc{L}_i^{\top} v_{q} -g_{q,i}(v_{d,i},v_{q,i},P_i,Q_i),
\end{align}
\end{subequations}
%
with the output load current modelled by the CPL representation given by $g_i = [g_{d,i} \ g_{q,i}]^{\top}$, where
\begin{subequations}
\begin{align}
g_{d,i}(v_{d,i},v_{q,i},P_i,Q_i)  &= \frac{2}{3} \left(\frac{v_{d,i}}{v_{d,i}^2+v_{q,i}^2} P_i + \frac{v_{q,i}}{v_{d,i}^2+v_{q,i}^2} Q_i\right) \\
g_{q,i}(v_{d,i},v_{q,i},P_i,Q_i)  &= \frac{2}{3} \left(\frac{v_{q,i}}{v_{d,i}^2+v_{q,i}^2} P_i - \frac{v_{d,i}}{v_{d,i}^2+v_{q,i}^2} Q_i\right) 
\end{align}
\end{subequations}
In the above, $C_i \in \mbb{R}_{>0}$ denotes the nodal capacitance of each inverter output filter, $P_i, Q_i \in \mbb{R}$ describe the local active and reactive power demand and $\mc{L}_i$ is the respective i$th$ column associated with the network Laplacian matrix. By $v_i = [v_{d,i} \ v_{q,i}]^{\top} \in \mbb{R}^{2}$ we represent the nodal voltage state, while the input of the system is the local injected current $I_{\ts{inj},i} =  [I_{\ts{inj},d} \ I_{\ts{inj},q}]^{\top} \in \mbb{R}^{2}$. A common requirement in the literature of power systems is the constraint of the output RMS voltage in a desired subset centred around the rated Microgrid voltage. The following assumption formalizes this, which will be useful in the closed-loop system analysis.

\begin{assumption}[RMS voltage constraint]\label{ass:constaint_set}
The network voltage constraint set is $\mbb{V} = \prod_{i\in \mc{M}} \mbb{V}_i$, where $\mbb{V}_i \coloneqq \left\{v_i \in \mbb{R}^2 \st \left|  \ \norm{v_i}-v_i^o \right| \leq v_i^{\ts{max}}  \right\}$, with $v_i^o > 0$ a rated RMS voltage and $0 < v_i^{\ts{max}} \ll v_i^o$ a maximum deviation from the rated value.
\end{assumption}
Following the above assumption, the state constraint is a Cartesian product of each local constraint set, indicating that there exist no coupled constraints in the network.  It is important to highlight at this point, that this work proposes a primary nodal controller. Requirements such as coupled constraints are commonly addressed in the secondary control level; the interested reader is referred to examples of relevant works in the literature \cite{agundis2018extended,arkhangelski2021day}. However, one of the novelties of the proposed controller is the ability to handle local state constrains within the primary control loop, by adopting a smooth feedback control law, as opposed to the commonly used reference saturation technique. This provides a straightforward technique to guarantee boundedness even during transient performance. 

\begin{remark}
We note that in the above model, we can also include the uncertainty arising by the intermittent nature of renewable power generation. Specifically, by implicitly assuming the existence of sufficient storage capacity at each DER unit such that the total power demand can always be satisfied, the input current is formulated as $ I_{\ts{inj},i} = I_{ren,i} + I_{\ts{stg},i}$, where $I_{ren,i} \in \mbb{R}^2$ and $I_{\ts{stg},i}\in \mbb{R}^2$ are the injected currents from the renewable source and storage unit respectively. Thus, the renewable unit acts as a constant power source, modelled as $I_{\ts{ren},i} = [g_{d,i}(v_i,P_{\ts{ren},i},Q_{\ts{ren},i}), \ g_{q,i}(v_i,P_{\ts{ren},i},Q_{\ts{ren},i})]^{\top}$ with $P_{\ts{ren},i}, Q_{\ts{ren},i}$ the supplied active and reactive power. Then, the subsequent analysis can be performed on the total power interaction $P_i = P_{\ts{ren},i} - P_{l,i}$ and $Q_i = Q_{\ts{ren},i} - Q_{l,i}$ with the load demand $(P_{l,i},Q_{l,i})$.
\end{remark}
\section{Problem Formulation}\label{sec:Problem_Form}
\subsection{Derivation of cascaded system dynamics}
In this section we formally present the investigated problem and formulate the control objectives. Firstly, it is noted that the system uncertainty arises from the perturbations of the CPL demand. Let the active and reactive power required by the load be a deviation from some known constant value, \ie
\begin{subequations}
\begin{align}
P_i &= \bar{P}_i + \delta P_i, \\
Q_i &= \bar{Q}_i + \delta Q_i, 
\end{align}
\end{subequations}
where $\bar{P}_i, \bar{Q}_i \in \mbb{R}_{>0}$ are known constant values and $\delta P_i, \delta{Q}_i \in \mbb{R}$ are the unknown deviations. We invoke the following assumption, establishing an upper bound on CPL variations.
\begin{assumption}\label{ass:load_demand}
For all $i \in \mc{M}$, there exist $\delta P_{\max,i} \in \mbb{R}_{>0}$ and $\delta Q_{\max,i} \in \mbb{R}_{>0}$ such that,
\begin{align*}
 \delta P_i  \in \mbb{W}_{P,i} &\coloneqq \left\{\delta P_i \in \mbb R \st \left|\delta P_i\right| \leq \delta P_{\max,i}\right\}, \\
\delta Q_i \in \mbb{W}_{Q,i} &\coloneqq \left\{\delta Q_i \in \mbb R \st \left|\delta Q_i \right| \leq \delta Q_{\max,i}\right\} 
\end{align*}
\end{assumption}
The aim of this study is to derive an analytic relation between the proposed controller parameters and the load demand. Then, the desired boundedness property is guaranteed by simply tuning the controller in accordance with the proposed methodology. In that sense, we require an assumption on a maximum load fluctuation in order to formulate a worst-case scenario condition. Following the above, it is possible to restructure the dynamical model (\ref{eq:True_System_Model_static}) to a pair of a nominal state, perceiving a constant disturbance, and an error describing the difference between the nominal and the true dynamics. To achieve this, we define a error state $e_i = [e_{d,i} \ e_{q,i}]^{\top}$ as
\begin{equation}
e_i =  v_i - z_i
\end{equation}
where  $z_i \coloneqq [z_{d,i} \ z_{q,i}]^{\top} \in \mbb{R}^2$ is a nominal state associated to a nominal subsystem with dynamic model given by

\begin{subequations}\label{eq:Nominal_Model}
\begin{align}
C_i \dot{z}_{d,i} = \bar{I}_{\ts{inj},di} + \omega_g C_i z_{q,i} -  \mc{L}_i^{\top} z_{d,i} - g_{d,i}(z_i, \bar{P}_i,\bar{Q}_i), \\
C_i \dot{z}_{q,i} = \bar{I}_{\ts{inj},qi} - \omega_g C_i z_{d,i} -  \mc{L}_i^{\top} z_{q,i} - g_{d,i}(z_i, \bar{P}_i,\bar{Q}_i).
\end{align}
\end{subequations}
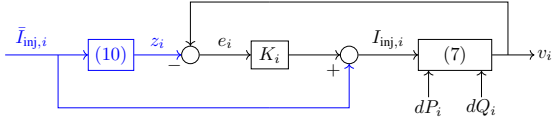
\begin{figure}[t]
	\centering
	\resizebox{0.9\columnwidth}{!}{%
		\tikzstyle{block} = [draw, rectangle, 
    minimum height=1em, minimum width=2em]
\tikzstyle{sum} = [draw, circle, node distance=1cm]
\tikzstyle{input} = [coordinate]
\tikzstyle{output} = [coordinate]
\tikzstyle{pinstyle} = [pin edge={to-,thin,black}]

\begin{tikzpicture}[auto, node distance=2cm]
    \node [output, name=input0] {};
    \node [input, name=input,right of=input0, node distance=1cm] {};
   	\node [block, right of=input,
            node distance=1cm,blue] (nominal) {$(10)$};    
    \node [sum, right of=nominal,node distance=1.5cm] (sum1) {};
    \node [block, right of=sum1,node distance=1.5cm] (K) {$K_i$};
    \node [sum, right of=K,node distance=1.5cm] (sum2) {};
     \node [block, right of=sum2,
            node distance=2cm] (system) {$\quad (7) \quad $};
     
     \node [output,above of=K,  node distance=1cm, name=f1] {};       
     \node [output,below of=K,  node distance=1cm, name=f2] {};

     \node at (8,-1) {$dP_i$}; 
   	 \node at (9,-1) {$dQ_i$}; 
 	  \node at (10.2,0) {$v_i$}; 
 	  
 	  \node at (3.2,-0.2) {$-$}; 
 	  \node at (6.2,-0.3) {$+$}; 

    \draw [->] (sum2) -- node[name=Inj] {$I_{\ts{inj},i}$} (system);
    \draw [->] (K) -- node[name=Ke] {} (sum2);
    \draw [->] (sum1) -- node[name=e] {$e_i$} (K);
    \draw [->,blue] (nominal) -- node[name=z] {$z_i$} (sum1);
    \draw [->,blue] (input) -- (nominal);
	\draw [-] (system) --++ (1,0) coordinate(v);
	\draw [->] (v) --++ (0.5,0) coordinate(out);
	\draw [-] (v) |- (f1);
	\draw [->] (f1) -| (sum1);
	\draw [-,blue] (input) |- (f2);
	\draw [->,blue] (f2) -| (sum2);
	\draw [-,blue] (input0) -- node[name=nominal_u] {$\bar{I}_{\ts{inj},i}$} (input);
	\draw [->] (8,-0.8)  --++ (0,0.5);
	\draw [->] (9,-0.8)  --++ (0,0.5);

\end{tikzpicture}
        }
        \caption{Control diagram of the proposed nodal control law. The nominal subsystem is depicted by blue colour.}
        \label{fig:C-diagram} 
\end{figure}
In the above, $\bar{I}_{\ts{inj},i} \in \mbb{R}^2$ is the nominal injected current \ie the control input of the nominal subsystem. Then, similarly to \cite{michos2023control}, by defining the injected current of the true voltage dynamics (\ref{eq:True_System_Model_static}) as a feedback control law of the form
\begin{equation}\label{eq:control_law}
I_{\ts{inj},i} = -\begin{bmatrix}
K_i && 0 \\
0 && K_i \\
\end{bmatrix} \begin{bmatrix}
e_{d,i} \\
e_{q,i}
\end{bmatrix} +  \bar{I}_{\ts{inj},i}
\end{equation}
the dynamic model of the error subsystem is computed
\begin{subequations}\label{eq:Error_Model}
\begin{align}
C_i \dot{e}_{d,i} &=  - K_i e_{d,i} +\omega_g C_i e_{q,i}-  \mc{L}_i^{\top} e_d - g_{d,i}(e_i+z_i,P_i,Q_i), \\
C_i \dot{e}_{q,i} &= - K_i e_{q,i}   -\omega_g C_i e_{d,i} -  \mc{L}_i^{\top} e_q - g_{d,i}(e_i+z_i,P_i,Q_i).
\end{align}
\end{subequations}
%
%
%
%
%
%
 The proposed control structure is illustrated in Fig. \ref{fig:C-diagram}. In order to simplify the adopted model, we have made the implicit assumption that the nominal voltage controller enforces $z_{q,i}=0$ at all times. This is a common strategy in the literature of power systems, where we require one component of the d-q model to be identically zero, see \cite{d2015virtual}. The following sections will analytically present how this assumption can be enforced. However, in the proposed control architecture, the nominal dynamics are uncertainty-free, \ie it is substantially easier to satisfy the invoked assumption from a control perspective.

\begin{remark}
It is important to highlight that (\ref{eq:control_law}) is a decentralized, primary level, control law. The control scheme requires local measurement of the output inverter voltage, in order to compute the error state using local variables and the nodal injected current, \ie it does not require knowledge of the neighbouring state nor line parameters.
\end{remark}

The original nodal capacitor dynamics have been restructured to a pair of d-q states described by (\ref{eq:Nominal_Model}) and (\ref{eq:Error_Model}) respectively. The aim of this paper is to design a unified control strategy, such that the nodal nominal solution operates as a reference trajectory for the true system, while the error is bounded in a compact set containing the origin. 

\subsection{Control Objectives}\label{ssection:Control_Obj}
The control objectives can be formulated as:
\begin{enumerate}
\item Guarantee the existence of a non-empty compact ``safe" set $\mc{S}_i \subset \mbb{R}^2$ containing the origin such that $e_i(0)\in \mc{S}_i$ implies $e_i(t) \in \mc{S}_i, \ \forall t>0$.
\item Achieve $\lim_{t \to \infty} z_i(t) = z_i^*$, where $z_i^*\in \mbb{R}^2$ is a desired reference vector for the nodal nominal trajectory.
\item Ensure that if $v_i(0)\in \mbb{V}_i$ then $v_i(t) \in \mbb{V}_i, \ \forall t>0$, where $\mbb{V}_i \subset \mbb{R}^2$ is a set with strictly positive elements and is centred around a common rated voltage value $v_i^o \in \mbb{R}^2$, where $v_i^o = v_j^o$ for all $(i,j) \in \mc{E}$.
\end{enumerate}

\section{Boundedness of error trajectory using a control barrier function}\label{sec:Error}

In this section, we aim to derive conditions on the magnitude of the feedback gain, such that the first control objective, defined in Section \ref{ssection:Control_Obj}, is satisfied. Essentially, this will enable a type of a ``tube" behaviour of the true voltage dynamics, where the distance between the true and nominal voltage trajectories determines the tube width. We are interested in the necessary conditions such that the tube width does not grow unbounded, which could possibly lead to unstable behaviour. Let $\mbb{D} \subset \mbb{R}^2$ and consider a continuously differentiable quadratic function $h \st \mbb{D} \to \mbb{R}$ given by,
\begin{equation}\label{eq:h}
h_i(e_i) = e_i^{\top} e_i.
\end{equation}
Furthermore, we define a continuously differentiable function $b_i \st \mbb{D} \to \mbb{R}$, given by $b_i(e_i) =  \bar{e} - h(e_i)$, where  $\bar{e} \in \mbb{R}_{>0}$ has strictly positive elements. This allows us to define a candidate positive invariant ``safe" set as the $\bar{e}$-level set of $h(\cdot)$ or equivalently,
\begin{subequations}\label{eq:error_PI_set}
\begin{align}
\mc{S}_i \coloneqq \left\{ e_i \in \mbb{R}^2 \st b_i(e_i) \geq 0   \right\}, \\ 
\partial \mc{S}_i \coloneqq \left\{ e_i \in \mbb{R}^2 \st  b_i(e_i) =  0  \right\}.
\end{align}
\end{subequations}
Our aim is to show that the system vector field of any point on the boundary of the safe set is pointing inwards, and thus at any time instant $\tau > 0$, the intersection of the tube in d-q coordinates is given by $\mc{S}_i \oplus \{z_i(\tau)\}$. This requirement can be formally stated by first defining the Lie derivative
\begin{equation}\label{eq:condition}
\mf{L}_{\dot{e}} b_i(e_i) \coloneqq -\frac{\partial h_i(e_i)}{\partial e_i} \frac{de_i}{dt}.
\end{equation}

The task now becomes formulating sufficient conditions on the control law such that (\ref{eq:condition}) is non-negative, \ie the control law enforces Positive Invariance of the ``safe set", defined by the CBF $b(\cdot)$, under the solution of the dynamics, for  any $\delta P_i  \in \mbb{W}_{P,i}$, and $\delta Q_i \in \mbb{W}_{Q,i}$. Since we are only interested in showing positive invariance only of $\mc{S}_i$, as opposed to enforcing this condition on every sublevel set, we can relax the condition to investigate the definiteness of $\inf_{e_i \in \partial \mc{S}_i} \left\{ \mf{L}_{\dot{e}} b_i(e_i)\right\}$. In other words, we investigate the behaviour of the network considering the states on the boundary of the set, where the respective nodes supply power to the network, \ie it holds that $\norm{e_i} \geq \norm{e_j}, \ \forall (i,j) \in \mc{E}$. Then, similarly to \cite{michos2023control}, exploiting this property decouples  $\inf_{e_i \in \partial \mc{S}_i} \left\{ \mf{L}_{\dot{e}} b_i(e_i)\right\}$ and allows the derivation of a lower bound such that
\begin{equation}\label{eq:V_f}
\inf_{e_i \in \partial \mc{S}_i} \left\{ \mf{L}_{\dot{e}} b_i(e_i)\right\} \geq  - \frac{\alpha_{nom,i}(e_i,z_{i})}{\alpha_{den,i}(e_i,z_{i})}.
\end{equation}
where,
 \begin{equation}\label{eq:V_tilde}
\begin{aligned}
\alpha_{nom,i}&(e_i,z_{i}) = e_{d,i}^4(-3 K_i z_{d,i}) + e_{d,i}^3(-6 K_i z_{d,i}^2+ 2\bar{P}_i)+ \\ 
& e_{d,i}^2e_{q,i}^2(-3 K_i z_{d,i}) + e_{d,i}^2 e_{q,i}(-3K_i z_{d,i}+2 \bar{Q}_i) +\\
& e_{d,i}^2 (-3 K_i z_{d,i}^3 +2\bar{P}_i z_{d,i} - 2 \delta P_i z_{d,i}) + \\ 
& e_{d,i} e_{q,i}^2(-6K_iz_{d,i}^2+2 \bar{P}_i) + e_{d,i}e_{q,i}(-4\bar{Q}_i z_{d,i}) +\\ 
& e_{q,i}^2(-3K_iz_{d,i}^3-2\bar{P}_iz_{d,i} - 2 \delta P_i z_{d,i})  +\\ 
& e_{q,i}^3(-3K_iz_{d,i} - 2 \bar{Q}_i) + e_{d,i} (-2\delta P_i z_{d,i}^2) + \\ 
& e_{q,i}(2\delta Q_i z_{d,i}^2).
\end{aligned}
\end{equation}
and
\begin{equation}\label{eq:a_den}
\alpha_{den,i} = z_{d,i} \left((e_{d,i} + z_{d,i})^2 +e_{d,i} \right)
\end{equation}
The following collection of results formulate sufficient conditions on the proposed control law (\ref{eq:control_law}), such that the right-hand-side of (\ref{eq:V_f}) is non-negative. The next Lemma investigates the definiteness of (\ref{eq:a_den}).
\begin{figure}[t]
    \centering
    \begin{subfigure}[t]{0.24\textwidth}
        \centering
        \includegraphics[scale=0.32]{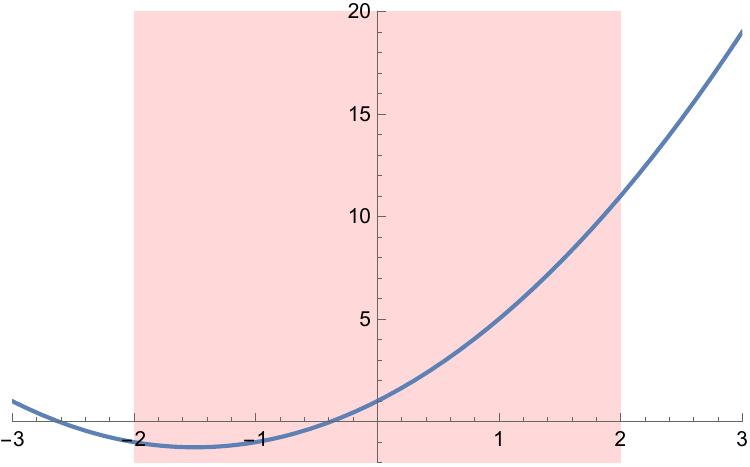}
        \caption{Case when $z_{d,i} < 3.414$}
    \end{subfigure}%
    ~ 
    \begin{subfigure}[t]{0.24\textwidth}
        \centering
        \includegraphics[scale=0.32]{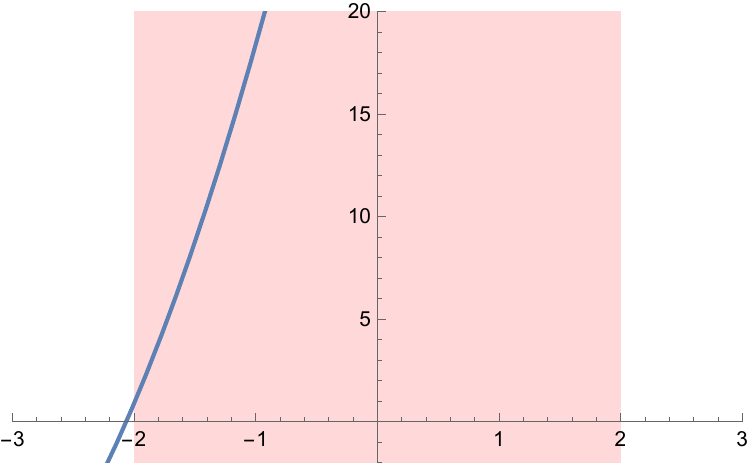}
        \caption{Case when $z_{d,i} > 3.414$}
    \end{subfigure}
    \caption{Numerical illustration of Lemma \ref{lem:zd_bound}, for $\bar{e} = 2 \ V$ with $\bar{e} + \sqrt{\bar{e}}=3.414 \ V$. Shaded region represents $|e| \leq \bar{e} \ V$. }
    \label{fig:lemma_1}
\end{figure}
\begin{lemma}\label{lem:zd_bound}
For all $i \in \mc{M}$, if the nominal state satisfies
\begin{equation}\label{eq:zd_bound}
z_{d,i} > \bar{e} + \sqrt{\bar{e}},
\end{equation}
then $\ts{Image}\left(\alpha_{den,i}\right) \subset \mbb{R}_{>0}$.
\end{lemma}
\begin{proof}
We will prove positive-definiteness of (\ref{eq:a_den}) directly from the properties of the function. First, we assume that $z_{d,i}>0$, \ie the d-component of the nominal voltage is  strictly positive at all times. Setting the first derivative of the function to zero yields the extreme point,
\[
\frac{\partial \alpha_{den,i}  }{\partial e_{d,i}} = 0 \Rightarrow e^*_{d,i} = - \frac{z_{d,i}^2+0.5}{z_{d,i}} 
\]
which is strictly negative since  $z_{d,i}>0$. Furthermore, it is straightforward to show that the function is convex and thus $e^*_{d,i}$ is a unique global minimizer. Assume, now, that (\ref{eq:a_den}) attains a negative value at the global minimum and thus obtains two real roots. In order to prove the Lemma statement, it is necessary to formulate a condition on the magnitude of the nominal voltage $z_{d,i}$, such that the largest real root of the function is smaller than the lower bound of the projection of $\mc{S}_i$ on d-axis. To formally state this, we require the largest root of $\frac{\alpha_{den,i}}{z_{d,i}}=0$ to satisfy
\[
\frac{-2 z_{d,i} +1 + \sqrt{4z_{d,i} +1 }}{2}  < -\bar{e},
\]
which yields,
\[
z_{d,i} > \bar{e} + \sqrt{\bar{e}}.
\]
\end{proof}
\begin{corollary}
If (\ref{eq:zd_bound}) holds, then the nodal error dynamics in (\ref{eq:Error_Model}) are Lipschitz continuous over $\mc{S}_i$.
\end{corollary}
The result of Lemma \ref{lem:zd_bound} is illustrated in Fig.~\ref{fig:lemma_1}. Setting the error bound at $\bar{e}=2 \ V$ as an example, we can compute the lower bound on the nominal voltage by Lemma \ref{lem:zd_bound} as $\bar{e} + \sqrt{\bar{e}}=3.414 \ V$. It is seen that (\ref{eq:a_den}) obtains strictly positive values in the desired region when the lower bound is satisfied and changes sign only when $z_{d,i} <3.414 \ V$. The next result establishes a sufficient condition such that (\ref{eq:V_tilde}) is a negative definite function.
\begin{lemma}\label{lem:nom_gain_bound}
If the feedback gain satisfies
\begin{equation}\label{eq:K_bound}
K_i \geq \beta_i (z_i)
\end{equation}
where $\beta_i \st \mbb{R}^2 \to \mbb{R}$,
\begin{multline}
\beta_i(z_i) \coloneqq \frac{2}{3} \frac{ \frac{1+z_{d,i}}{z_{d,i}} \bar{P}_i  +2\bar{Q}_i + 2\frac{\bar{e}+z_{d,i}}{\bar{e}}\delta P_{\max,i} +  \frac{z_{d,i}}{\bar{e}}\delta Q_{\max,i}}{(\bar{e}-z_{d,i})^2 - \bar{e}},
\end{multline}
then for all $e_{i} \in \partial \mc{S}_i $ and $i \in \mc{M}$ it holds that $\ts{Image}\left(\alpha_{nom,i}(e_i,z_{i})\right) \subset \mbb{R}_{<0}$.
\end{lemma}
\begin{proof}
We begin this proof by rearranging (\ref{eq:V_tilde}) as,
\begin{multline}\label{eq:a_nom_v2}
\alpha_{nom,i}(e_i,z_{i}) = -3K_i f_{K}(e_i) + 2\bar{P}_i f_{\bar{P}}(e_i) +2\bar{Q}_if_{\bar{Q}}(e_i) \\
- 2\delta P_i f_{\delta P}(e_i)+ 2\delta Q_i f_{\delta Q}(e_i)
\end{multline}
where,
\begin{subequations}
\begin{align}
f_{K}(e_i) =&  z_{d,i} e_{q,i} \Big( e_{q,i}^2 + e_{q,i} \big( e_{d,i} + z_{d,i}\big)^2 + e_{d,i}^2\Big) \nonumber \\ 
 &+  z_{d,i} e_{d,i}^2 \left(e_{d,i} + z_{d,i}\right)^2,  \\
f_{\bar{P}}(e_i) =& e_{d,i}^3 + e_{d,i}^2 z_{d,i} + e_{d,i} e_{q,i}^2 - z_{d,i} e_{q,i}^2, \\
f_{\bar{Q}}(e_i)  =& e_{d,i}^2 e_{q,i} - 2 z_{d,i}e_{d,i}e_{q,i} - e_{q,i}^3, \\
f_{\delta P}(e_i) =& z_{d,i} e_{d,i}^2 + z_{d,i}e_{q,i} ^2 + z_{d,i}^2 e_{d,i}, \\
f_{\delta Q}(e_i) =& e_{q,i} z_{d,i}^2.
\end{align}
\end{subequations}
Investigating the properties of $f_{K}(\cdot)$, it is straightforward to show that for $e_i \in \partial \mc{S}_i$ the function obtains strictly positive values. This follows from the fact that in the only case where  $f_{K}(\cdot)$ is not a sum of strictly positive functions is when $e_{q,i} = -\bar{e}$. However, utilizing Lemma \ref{lem:zd_bound}, the second summand of the first term remains strictly positive and dominates the other two terms inside the parenthesis. Thus, a sufficient condition for (\ref{eq:a_nom_v2}) to be strictly negative is computed as
\begin{align*}
K_i \geq& \frac{2}{3\min_{e_i \in \partial \mc{S}_i} \{f_{K}(e_i)\}} \Bigg( \max_{e_i \in \partial \mc{S}_i} \Big\{\bar{P}_i f_{\bar{P}}(e_i) \Big\} \\
& + \   \max_{e_i \in \partial \mc{S}_i}\Big\{\bar{Q}_i f_{\bar{Q}}(e_i)\Big\} + \max_{e_i \in \partial \mc{S}_i}\Big\{\delta P_{max,i} f_{\delta P}(e_i)\Big\} \\
& +  \max_{e_i \in \partial \mc{S}_i} \Big\{\delta Q_{\max,i} f_{\delta Q}(e_i)\Big\} \Bigg)
\end{align*}
Note that the denominator of the right-hand-side only vanishes at $(0,0) \notin \partial S_i$, which implies that the bounding function is continuously differentiable on $\partial S_i$. In addition, the above satisfy both conditions for the application of the Extreme Value Theorem; the set $\partial \mc{S}_i$ is compact and each individual function is continuous on this set. Therefore, there always exist at least one solution to each optimization problem. Applying the Lagrange method of multipliers to each individual optimization problem allows the computation of the bounding function
\begin{align*}
K_i \geq \frac{2}{3} \frac{ \frac{1+z_{d,i}}{z_{d,i}} \bar{P}_i  +2\bar{Q}_i + 2\frac{\bar{e}+z_{d,i}}{\bar{e}}\delta P_{\max,i} +  \frac{z_{d,i}}{\bar{e}}\delta Q_{\max,i}}{(\bar{e}-z_{d,i})^2 - \bar{e}}
\end{align*}
Setting the right hand side to $\beta_{i}(z_i)$ completes the proof.
\end{proof}
\begin{remark}
Lemma \ref{lem:zd_bound} can be used to conclude that the lower bound on $K_i$ is strictly positive. This follows from the fact that the nominator is a sum of strictly positive functions. Then, applying the condition of Lemma \ref{lem:zd_bound}, it can be directly shown that the denominator attains strictly positive values.
\end{remark}
We are now in position to establish positive invariance of (\ref{eq:error_PI_set}) under the solution of the error dynamics. This property is proven in the next result.
\begin{figure}[t]
	\centering
	\resizebox{0.4\columnwidth}{!}{%
		\includegraphics[scale=0.8]{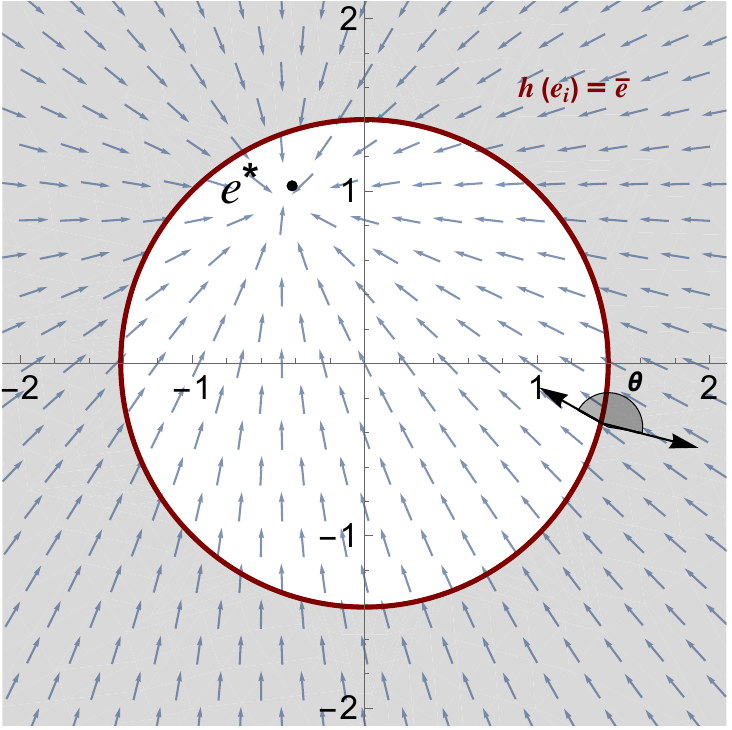}
        }
        \caption{Illustration of Prop. 1. The ``unsafe" region is depicted by light grey. } 
        \label{fig:PI_set} 
\end{figure}
\begin{proposition}[Boundedness of Error Dynamics]\label{prop:PI_set}
If the feedback gain of the control law in (\ref{eq:control_law}) satisfies (\ref{eq:K_bound}) and the initial condition of the error dynamics in (\ref{eq:error_PI_set}) satisfies $e(0) \in \mc{S}_i$ with set $\mc{S}_i$ defined in (\ref{eq:error_PI_set}), then $b\st \mbb{D} \to \mbb{R}$ is a control barrier function with
\begin{equation}
\inf_{e_i \in \partial \mc{S}_i} \left\{ \mf{L}_{\dot{e}} b_i(e_i)\right\} \geq  0
\end{equation}
and the nodal error trajectories satisfy $e_i(t) \in \mc{S}_i$ for all $t \geq 0$ and $i \in \mc{M}$.
\end{proposition}
\begin{proof}
By construction, the set $\partial \mc{S}_i = h^{-1}\left(\{\bar{e}\}\right)$ is closed, since it is defined as the pre-image of the closed  set $\{\bar{e}\}$ and $h \st \mbb{R}^n \to \mbb{R}$ is a continuously differentiable function. In addition, Lemma \ref{lem:zd_bound} guarantees Lipschitz continuity of the nodal error dynamics (\ref{eq:Error_Model}) over $\mc{S}_i$. Considering Lemma \ref{lem:zd_bound} and Lemma \ref{lem:nom_gain_bound} we deduce
\begin{equation}
\inf_{e_i \in \partial \mc{S}_i} \left\{ \mf{L}_{\dot{e}} b_i(e_i)\right\} \geq  - \frac{\alpha_{nom,i}(e_i,z_{i})}{\alpha_{den,i}(e_i,z_{i})}  > 0, \  \forall i \in \mc{M}.
\end{equation}
which implies that $b(\cdot)$ is a control barrier function. Furthermore, the inner product between the error subsystem vectorfield and the respective normal vector satisfies
\begin{align}
\innerproduct{\nabla h_i}{\dot{e}_i} \vert_{e_i \in \partial \mc{S}_i} &= -\mf{L}_{\dot{e}} b_i(e_i) \leq -\inf_{e_i \in \partial \mc{S}_i} \left\{ \mf{L}_{\dot{e}} b_i(e_i)\right\} \leq 0
\end{align}
Therefore, all conditions of the Bony-Brezis Theorem are satisfied and the set $\partial \mc{S}_i$, and by induction $\mc{S}_i$, is positive invariant under the solution of the dynamics. This implies that any nodal trajectory starting at time $t=0$ satisfies $e_i(0) \in \mc{S}_i$, then $e_i(t) \in \mc{S}_i, \forall t \geq 0$. This concludes the proof.
\end{proof}
The result of Prop. \ref{prop:PI_set} is illustrated in Fig.~\ref{fig:PI_set}. For $\theta$ denoting the angle between the normal vector on $\partial \mc{S}$ and the system vector field at some point  $e_i \in \partial \mc{S}$, the derived conditions on $z_{d,i}$ and $K_i$ enforce $|\theta| > \frac{\pi}{2}$, \ie $\mc{S}_i$ is a robust positive invariant set, \ie it is positive invariant for all disturbance perturbations. Satisfying the conditions on the nominal voltage and the feedback gain derived in Lemma \ref{lem:zd_bound} and Lemma \ref{lem:nom_gain_bound} respectively, enforces positive invariance of the level-set $\mc{S}_i$ under the solution of the nodal error dynamics.

\begin{remark}
A detail that is revealed within the provided analysis on the proposed control law (\ref{eq:control_law}), is that smaller values on the bound of the state error $\bar{e}$ result in a larger lower bound on the feedback gain $K_i$ in (\ref{eq:K_bound}). This in turn implies more aggressive control inputs, which, in this case study, is the injected current in (\ref{eq:control_law}). This could potentially collide with a common requirement in the literature of power systems, which is inverter current limitation \cite{rocabert2012control}. However, the provided analysis on the closed-loop dynamics quantifies specific tuning guidelines to help accommodate actuation limits. Specifically, for a given bound on the power demand, the invariance analysis indirectly guarantees a tunable uniform bound on the injected current. This is done via tuning the width of the tube, \ie the magnitude of true nodal voltage fluctuations quantified by $\bar{e}$, where larger values relaxes the bound on the required injected current, since part of the CPL is satisfied via deviations of the true voltage from the nominal value.
\end{remark}
Following the guidelines formulated in this section, guarantees that the first control objective is satisfied. The next sections are associated with the design of a constrained control architecture for the nominal dynamics that satisfy Lemma \ref{lem:zd_bound} and incorporate the desired operational constraints within the control designed process.
\section{Constrained regulation of the nominal voltage dynamics}\label{sec:Nominal}

In this section, a constrained control strategy is developed for the nominal dynamics. In order to satisfy the remaining control objectives, the proposed nominal controller is required to enforce a ``tightened" bound on the nominal nodal trajectory such that the true state is enclosed within the desired bound at all times. Then, it is also desired to achieve fast regulation to a desired reference point and guarantee stability of the closed-loop dynamics in an analytic fashion. Considering Assumption \ref{ass:constaint_set} and the results of the previous section,  it is necessary to satisfy
\begin{equation}\label{eq:set_equation}
\mc{S}_i \oplus \mc{Z}_i \subseteq \mbb{V}_i,
\end{equation}
where  $\mc{Z}_i \subset \mbb{R}^2$ is the constraint set for the nominal subsystem.  Considering Assumption \ref{ass:constaint_set}, $\mc{Z}_i$ takes the form
\begin{equation}
\mc{Z}_i \coloneqq \left\{ z_i \in \mbb{R}^2 \st z_i^o - \gamma \leq \sqrt{z_{d,i}^2 + z_{q,i}^2} \leq  z_i^o + \gamma \right\},
\end{equation}
where $\gamma>0$ is a constant characterizing the maximum allowed deviation from the Microgrid rated RMS voltage $z_i^o \in \mbb{R}$. Then, according to the third control objective we set $z_i^o = v_i^o$. To facilitate the above requirement, we introduce a coordinate shift of the nominal voltage dynamics as a deviation from the rated value, \ie $\tilde{z}_i = z_i-z_i^o$ and $\tilde{I}_{\ts{inj},i} = \bar{I}_{\ts{inj},i} - I_i^o$. This yields the shifted nominal subsystem
\begin{subequations}\label{eq:Nominal_Model_Shifted}
\begin{align}
C_i \dot{\tilde{z}}_{d,i} = \tilde{I}_{\ts{inj},di} +I_{di}^o + \omega_g C_i & \tilde{z}_{q,i} -  \mc{L}_i^{\top} \tilde{z}_{d,i} \nonumber \\ 
&- g_{d,i}(\tilde{z}_i+z_i^o, \bar{P}_i,\bar{Q}_i), \\
C_i \dot{\tilde{z}}_{q,i} = \tilde{I}_{\ts{inj},qi} +I_{qi}^o- \omega_g C_i & \tilde{z}_{d,i} -  \mc{L}_i^{\top} \tilde{z}_{q,i} \nonumber \\
&- g_{q,i}(\tilde{z}_i+z_i^o, \bar{P}_i,\bar{Q}_i).
\end{align}
\end{subequations}
%
%
We note that $I_i^o$ results from computing the equilibrium map of the nominal subsystem and substituting for the rated voltage. Therefore, the problem now becomes bounding the shifted nominal dynamics around the origin. To this aim, we propose the nominal control law defined as $\tilde{I}_{\ts{inj},i} \st \mbb{R}^2 \times \mbb{R}^2  \to \mbb{R}^2$, given by
\begin{multline}\label{eq:nominal_control_law}
\tilde{I}_{\ts{inj},i} = \begin{bmatrix}
- K_{d,i} && 0 \\
0 && - K_{q,i}
\end{bmatrix}
\begin{bmatrix}
\tilde{z}_{d,i} \\
\tilde{z}_{q,i}
\end{bmatrix}
+
\begin{bmatrix}
M_i && 0 \\
0 && 1
\end{bmatrix}
\begin{bmatrix}
\sigma_{d,i} \\
\sigma_{q,i}
\end{bmatrix} \\
+\begin{bmatrix}
0 \\
\omega_g C_i  \left(\tilde{z}_{q,i} + z_{q,i}^o \right) - g_{q,i}(\tilde{z}_i + z_i^o, \bar{P}_i,\bar{Q}_i).
\end{bmatrix}
\end{multline}
where $\left[ \sigma_{d,i} \ \sigma_{q,i}\right] \in \mbb{R}^2$ are the respective d-q integrator states with dynamics given by
\begin{subequations}\label{eq:nominal_control_law_sigma}
\begin{align}
\dot \sigma_{d,i} &= k_{Id,i}(1-\sigma_{d,i}^2)(\hat{z}_{d,i} - \tilde{z}_{d,i}), \label{eq:integrator_d} \\
\dot \sigma_{q,i} &= -k_{Iq,i} \tilde{z}_{q,i}. \label{eq:integrator_q} 
\end{align}
\end{subequations}
In the above $K_{d,i},k_{Id,i},M_i,K_{q,i},k_{Iq,i} \in \mbb{R}_{>0}$ are the controller tuning parameters for the respective d-q components, while $\hat{z}_{d,i} \in \mbb{R}$ denotes the desired shifted reference point.  Note that (\ref{eq:nominal_control_law}), (\ref{eq:nominal_control_law_sigma}) together with (\ref{eq:control_law}), result in a smooth feedback control law, that, as the following results will demonstrate, enforces boundedness of the state trajectories. Therefore,  the proposed approach overcomes the instability problems that are associated with non-smoothness of the closed loop dynamics, e.g. by using saturation devices \cite{xin2016synchronous}. 

The remaining of this section is dedicated in providing the necessary theoretical analysis such that the adopted control law guarantees boundedness and stability of the nominal dynamics. The control law in (\ref{eq:nominal_control_law}) decouples the q-component of the nominal state from the d-component nominal dynamics. It is important to highlight that the adopted methodology allows such an assumption without requiring knowledge of the load demand, since, in the nominal setting, this is constant to a predefined value. Therefore, the load demand remains an unknown system disturbance in the original problem setting. The following result investigates the stability of the system equilibria and derives necessary conditions on the tuning parameters to counteract the destabilizing effect of the CPL.
\begin{proposition}[Stability of Nominal Dynamics]\label{prop:Nominal_Stability}
Consider the nodal nominal dynamics given by (\ref{eq:Nominal_Model}), (\ref{eq:nominal_control_law}) and (\ref{eq:nominal_control_law_sigma}) and an equilibrium point $\hat{x}_i = (\hat{z}_{d,i},\hat{\sigma}_{d,i},0,0)$ , if $\hat{\sigma}_{d,i} \in (-1,1)$ and
\begin{equation}\label{eq:K_condition}
K_{d,i} > \frac{2}{3}\frac{\bar{P}_i }{\hat{z}^2_{d,i}},
\end{equation}
then the closed-loop nominal system (\ref{eq:Nominal_Model_Shifted}) is asymptotically stable with respect to $\hat{x}_i$, for all $i \in \mc{M}$.
\end{proposition}
\begin{proof}
In order to derive asymptotic stability of the system equilibria, we study the network closed-loop nominal subsystem. With a slight abuse of notation these are given below,
\begin{subequations}
\begin{align}
[C] \dot{\tilde{z}}_d =& I_d^o -\left([K_{d}]+\mc{L}\right)\tilde{z}_d + [M]\sigma_d + [\omega_g C] \left(\tilde{z}_q + z_{q}^o \right)  \nonumber \\
& \mkern+180mu + \left[g_d\left(\tilde{z} + z^o,\bar{P},\bar{Q}\right)\right], \\
\dot{\sigma_d} =& [k_{Id}] \left[ 1-\sigma_d^2 \right] \left(\hat{z}_d - \tilde{z}_d\right), \\
[C]\dot{\tilde{z}}_q =& I_q^o-\left(\left[K_q\right]+\mc{L}\right) \tilde{z}_q + \sigma_q, \\
\dot{\sigma}_q =& \left[k_{Iq}\right] \tilde{z}_q. 
\end{align}
\end{subequations}
Considering an network equilibrium $\hat{x}$, the resulting Jacobian matrix of the system is computed as
\begin{equation}\label{eq:Jacobian}
\mc{J}=
\begin{bNiceArray}{cc|cc}[margin]
  \Block{2-2}<\normalsize>{\mc{J}_{11}} & & \left[\omega_g - \frac{2}{3\hat{z}_{d}^2} \bar{Q}\right] & \mathbf{0}_{n} \\
  & & \mathbf{0}_{n} & \mathbf{0}_{n} \\
  \hline
  \Block{2-2}<\normalsize>{\mathbf{0}_{2n}} & & \Block{2-2}<\normalsize>{\mc{J}_{22}}\\
  & & &
\end{bNiceArray}
\end{equation}
where 
\begin{align}
\mc{J}_{11} &= 
\begin{bmatrix}
[C]^{-1}\left(-[K_d] - \mc{L} + \left[\frac{2}{3 \hat{z}_d^2} \bar{P}\right] \right) && [C]^{-1}[M]  \\
-[k_{Id}] \left[1-\hat{\sigma}_d^2\right] && \mathbf{0}_{n \times n}
\end{bmatrix},\\
\mc{J}_{22}&=
\begin{bmatrix}
 -[K_q]-\mc{L} && \mathbf{\ts{I}}_{n} \\
 -[k_{Iq}] && \mathbf{0}_{n}
\end{bmatrix}.
\end{align}
Matrix (\ref{eq:Jacobian}) is a block triangular matrix, where it holds that $\ts{det}\left(\lambda \mathbf{\ts{I}_n} - J\right) = \ts{det}\left(\lambda \mathbf{\ts{I}_n} - J_{11}\right)\ts{det}\left(\lambda \mathbf{\ts{I}_n} - J_{22}\right)$, \ie, the eigenvalues are a combination of the block matrices on the diagonal \cite{axler2024linear}. Therefore, if both $\mc{J}_{11}$ and $\mc{J}_{22}$ are Hurwitz then $\mc{J}$ is also a Hurwitz matrix. Considering $\mc{J}_{11}$, the characteristic polynomial is computed
\begin{equation}\label{eq:char_J11}
\ts{det} \left( \lambda\mathbf{\ts{I}_n} - \mc{J}_{11} \right) = \left| \lambda^2 \mathbf{\ts{I}_n} + \lambda A + B\right| = 0, 
\end{equation} 
where
\begin{align}
A &=  [C]^{-1}\left([K_d] + \mc{L} - \left[\frac{2}{3 \hat{z}_d^2} \bar{P}\right] \right), \\
B &=  [C]^{-1} [M] [k_{Id}] \left[1-\hat{\sigma}_d^2\right].
\end{align}
Right multiplying (\ref{eq:char_J11}) by $\left|[C]\right|$ yields
\begin{equation}
\left| [C] \lambda^2 \mathbf{\ts{I}_n} + \lambda \mathbf{A} + \mathbf{B} \right| = 0,
\end{equation} 
where $\mathbf{A} = [C] A$ and $\mathbf{B} = [C]B$. In the above, each coefficient is a symmetric matrix and thus the expression defines a Quadratic Eigenvalue Problem (QEP). According to the QEP theory \cite{tisseur2001quadratic}, the eigenvalues are negative if each coefficient is positive definite. By construction, $\mathbf{B}$ is a diagonal and positive definite matrix and it also holds that $C \succ 0$, thus we focus our attention on the second coefficient. By the properties of a Laplacian matrix, it holds that the respective eigenvalues satisfy $\lambda_1=0 < \lambda_2 \leq \dots \leq \lambda_n$, and thus $\lambda_{\min} (-\mc{L}) = 0$. Considering that the rest of the summands of $\mathbf{A}$ are diagonal matrices, the smallest eigenvalue $\lambda_{\max} (\mathbf{A})$, corresponding to some i$th$ node, is given by
\begin{equation}
\lambda_{\max} (\mathbf{A}) = K_{d,i} - \frac{2}{3} \frac{\bar{P}_i}{\hat{z}_{d,i}^2}.
\end{equation}
Therefore, the condition $\lambda_{\max} (\mathbf{A}) <0$ yields
\begin{equation}
K_{d,i} > \frac{2}{3} \frac{\bar{P}_i}{\hat{z}_{d,i}^2}.
\end{equation}
Satisfying the above implies that $\mc{J}_{11}$ is a Hurwitz matrix. A similar methodology can be used to prove $\mc{J}_{22}$ is also Hurwitz under the condition that $K_{q,i} >0$ holds for all $i \in \mc{M}$. Therefore $\mc{J}$ is also Hurwitz, since its eigenvalues are given as a combination of $\mc{J}_{11}$ and $\mc{J}_{22}$. Therefore, the closed-loop system admits asymptotically stable equilibria in $\mc{Z}_i$. This concludes the proof.
\end{proof}
The closed-loop nominal subsystem admits asymptotically stable equilibria as long as condition (\ref{eq:K_condition}) holds. Nevertheless, on of the control objectives is to guarantee that the local bus voltage adheres to desired operational constraints. Prop. \ref{prop:Nominal_Stability} guarantees that there exist a local Lyapunov function and thus every subset of the respective domain is positive invariant; However, this result holds only locally in a neighbourhood of the respective equilibrium point.  Therefore, the above are not sufficient to guarantee positive invariance of $\mc{Z}_i$ under the solution of the nominal nodal dynamics. The desired property is proven in the following result.
\begin{proposition}[Boundedness of Nominal Voltage]\label{prop:Nominal_Boundedness}
Consider a shifted nominal voltage constraint set defined as 

\begin{equation}
\tilde{\mc{Z}}_i \coloneqq \left\{ \tilde{z}_i \in \mbb{R}^2 \st \tilde{z}_{q,i} = 0, \  -\delta_i - \tilde{z}_{\ts{m},i} \leq \tilde{z}_{d,i} \leq \tilde{z}_{\ts{m},i} \right\},
\end{equation}
where $\delta_i >0$ and $\tilde{z}_{\ts{m},i} = \frac{M_i}{K_{d,i}}$. If at time $t=0$ the initial conditions satisfy $z_i(0) \in \mc{\tilde{Z}}_i$ and $\sigma_{d,i}(0) \in (-1,1)$, and
\begin{equation}
K_{d,i} \geq \frac{2}{3} \frac{\bar{P}_i}{\delta_i \left(z^o_i - \tilde{z}_{\ts{m},i} -\delta_i \right) },
\end{equation}
holds for all $i \in \mc{M}$, then for all $t>0$, the nominal system states satisfy $\tilde{z}_i(t) \in \tilde{\mc{Z}}_i$ and $\sigma_{d,i}(t) \in (-1,1)$ .
\end{proposition}
\begin{proof}
The proof of this result will be provided in two steps, showing first boundedness of the d-component integrator dynamics and using this property to prove boundedness of the nominal voltages. The first part will be proven by contradiction. Consider the continuously differentiable integrator dynamics from (\ref{eq:integrator_d}). Let there exist a time instant $\tau_2$ such that the respective solution satisfies $|\sigma_{d,i}(\tau_2)| >1$. Continuity of the dynamics imply that there exists a time instant $\tau_1$ such that $|\sigma_{d,i}(\tau_1)| = 1$, and, furthermore, $\theta_1,\theta_2>0$ such that $|\sigma_{d,i}(\tau_1-\theta_1)| < 1$ and  $|\sigma_{d,i}(\tau_1+\theta_2)| > 1$, \ie the solution enters and leaves the set boundary defined at  $|\sigma_{d,i}(t)|= 1$. Nevertheless, the point $|\sigma_{d,i}(t)|= 1$ is an equilibrium point of the integrator dynamics, attracting or repelling the respective trajectories. This leads to a contradiction and thus for every possible trajectory with initial condition $\sigma_{d,i}(0) \in (-1,1)$ it holds that $\sigma_{d,i}(t) \leq (-1,1), \ \forall t \geq 0$.

In order to prove boundedness of the nominal voltage, we first note that Prop. \ref{prop:Nominal_Stability} implies that $\tilde{z}_{q,i}(t) =0$, for all $t>0$, since $\tilde{z}_{q,i}(0) =0$. This creates a one dimensional constraint set on the d-component of the nominal voltage, that, by construction, is closed and bounded. Similarly to the previous section, we will show set invariance using the inner product of the system vector field and the outside normal vector on the boundary of the constraint set. In regards to the upper bound this yields,
\begin{align*}
\left. \left \langle \dot{\tilde{z}}_{d,i}, 1 \right \rangle \right |_{\tilde{z}_{\ts{m},i}} =& - M_i + M_i \sigma_{d,i} - \frac{\bar{P}_i}{\tilde{z}_{\ts{m},i} + z_i^o} \leq 0
\end{align*}
Similarly, the inner product at the lower boundary is computed as 
\begin{align*}
\left. \left \langle \dot{\tilde{z}}_{d,i}, -1 \right \rangle \right |_{-\tilde{z}_{\ts{m},i}-\delta_i } =  -K_{d,i}\delta_i + \frac{2}{3} \frac{\bar{P}_i}{-\tilde{z}_{\ts{m},i} - \delta_i + z_i^o}.
\end{align*}
In order for the inner product to be negative, we deduce the condition
\begin{equation}\label{eq:K_d_condition_proof}
K_{d,i} \geq \frac{2}{3}\frac{\bar{P}_i}{\delta_i\left(z_i^o -\tilde{z}_{\ts{m},i}- \delta_i\right)}.
\end{equation}
Therefore, both conditions of the Bony-Brezis Theorem are satisfied; \ie the system solutions are unique and continuous and the inner product on the boundary of the desired set is non-positive. Therefore, we can conclude that $\mc{\tilde{Z}}_i$ is positive invariant under the solution of the nominal nodal dynamics and by direct extension $\mc{\tilde{Z}}$ is positive invariant with respect to the network nominal dynamics.
\end{proof}

\begin{corollary}\label{cor:nom_bound}
For $-\gamma \leq -\delta_i - \tilde{z}_{\ts{m},i}$ and $\tilde{z}_i(t=0) \in \tilde{\mc{Z}}_i$, it holds that $\tilde{\mc{Z}}_i \oplus \{z_i^o\} \subseteq \mc{Z}_i$ and $\mc{Z}_i$ is positive invariant under the solution of (\ref{eq:Nominal_Model}) with (\ref{eq:nominal_control_law}) for all $i \in \mc{M}$ and $t>0$.
\end{corollary}
The above results also allows us to draw conclusions on the true system trajectories in regards to the third control objective in Section \ref{ssection:Control_Obj}.
\begin{corollary}\label{cor:voltage_bound}
If $v_i(0) \in \mbb{V}_i$, then $v_i(t)\in \mbb{V}_i$ for all $t>0$ and $i \in \mc{M}$, with $v_i^{\ts{max}} = \sqrt{\bar{e}} + \gamma_i$.
\end{corollary}
\begin{proof}
By the triangular property of the norm function, it holds that
\begin{align}
    \norm{z_i} -\norm{e_i} \leq \norm{v_i} = \norm{z_i + e_i} \leq \norm{z_i} + \norm{e_i}.
\end{align}
By Prop. \ref{prop:PI_set} and \ref{prop:Nominal_Boundedness}, it holds that $\norm{e_i} \leq \sqrt{\bar{e}}$ and $z_i^o - \gamma_i \leq \norm{z_i} \leq z_i^o + \gamma_i$. Thus,
\begin{equation}
     \left(z_i^o - \gamma_i\right)-\sqrt{\bar{e}} \leq \norm{v_i} \leq \sqrt{\bar{e}} + (z_i^o + \gamma_i).
\end{equation}
Since $z_i^o = v_i^o$, it holds that $\abs{\norm{v_i} - v_i^o} \leq  \sqrt{\bar{e}} + \gamma_i \coloneqq v_i^{\ts{max}}$ for all $i \in \mc{M}$. This completes the proof.
\end{proof}

Having established the desired theoretic guarantees, we can now substitute $\bar{I}_{\ts{inj},i} = \tilde{I}_{i} - \bar{I}_{o,i}$ and (\ref{eq:nominal_control_law}) to (\ref{eq:control_law}), in order to derive the final version of the proposed control scheme. In the case of a non-constant nominal voltage, the application of the proposed control law requires knowledge of the nominal trajectory, \ie $z_i(t)$. If the nominal voltage references are known a-priori, which is common in power systems due to day-ahead scheduling, then the computation of $z_i(t)$ can be performed offline. Otherwise there is a need to compute this online by some additional steps. In other words, we require as input the solution of a distributed system to the proposed decentralized control law. Nevertheless, a common requirement in AC Microgrids is keeping the nodal nominal voltage constant close to some desired value. In this scenario, it is straightforward to compute $z_i(t)$ and $\bar{I}_{\ts{inj},i}$ offline by a direct substitution of the constant value, hence facilitating a completely decentralized implementation. 
\begin{remark}
Considering the theoretic results of this section, the feedback gain $K_{d,i}$ needs to be chosen to satisfy both conditions of Prop. \ref{prop:Nominal_Stability} and \ref{prop:Nominal_Boundedness}, \ie
\begin{equation}\label{eq:K_d_condition}
K_{d,i} \geq \frac{2}{3} \max \left\{\frac{\bar{P}_i}{\left(- \tilde{z}_{\ts{m},i} -\delta_i \right)^2}, \ \frac{\bar{P}_i}{\delta_i \left(z^o_i -\tilde{z}_{\ts{m},i} -\delta_i \right) } \right \}
\end{equation}
\end{remark}
\section{Stability of the Cascaded System}\label{sec:Stability}

By collecting the results of the previous sections, the original system dynamics in (\ref{eq:True_System_Model_static}) are now represented by the equivalent system model described by (\ref{eq:Error_Model}), (\ref{eq:Nominal_Model_Shifted}), (\ref{eq:nominal_control_law}), (\ref{eq:nominal_control_law_sigma}). The new cascaded structure describes the true voltage trajectory $v(t)$ as a summation of the respective error and nominal trajectories, \ie  $v(t) = e(t) + z(t) = e(t) + \tilde{z}(t) + z^o$. The last requirement is combining the theoretic results of the previous sections to guarantee asymptotic stability of the cascaded dynamics. Due to the presence of the non-constant load demand, it is required to describe the stability properties with respect to an equilibrium \textit{set} rather than an equilibrium \textit{point}. To this end, we define an equilibrium set for the cascaded network dynamics $\Omega \subseteq \mbb{R}^{6n}$ given by
\begin{multline}\label{eq:Omega_set}
\Omega \coloneqq \Big\{\left(e,\tilde{z},\sigma_d,\sigma_q \right) \in \mbb{R}^{6n} \st \\
e \in \mc{S}, \ \tilde{z} = \left(\hat{z}_d,0\right)^{\top}, \sigma_d =\hat{\sigma}_d, \ \sigma_q = 0  \Big\},
\end{multline}
where $\hat{\sigma}_d = [M]^{-1} \left([K_d] + \mc{L} \right)^{-1}\hat{z}_d -g_d\left(\hat{z}+z^o,\bar{P},\bar{Q}\right)$.
The asymptotic stability of the network dynamics with respect to $\Omega$ is proven in the next result.

\begin{theorem}[Stability of Network Dynamics]
Let $e_i(0) \in \mc{S}_i$, $\tilde{z}_i(0) \in \tilde{\mc{Z}}_i $, $\sigma_d(0) \in (-1,1) $, $\sigma_q(0)=0$ for all $i \in \mc{M}$. The network dynamics described by (\ref{eq:Error_Model}), (\ref{eq:Nominal_Model_Shifted}), (\ref{eq:nominal_control_law}), (\ref{eq:nominal_control_law_sigma}) are asymptotically stable with respect to the equilibrium set $\Omega$ and $\lim_{t \to \infty} v(t) = z^o + \left(\hat{z},0\right)^{\top} + e(t) $, with $e(t) \in \mc{S}$.
\end{theorem}
\begin{proof}
Leveraging the results of Prop. \ref{prop:Nominal_Stability}, the nominal dynamics (\ref{eq:Nominal_Model_Shifted}) admit a single asymptotically stable equilibrium in $\tilde{\mc{Z}}_i$ and $\tilde{\mc{Z}}_i$ is positive invariant under the solution of the nominal subsystem for all $i \in \mc{M}$. As a result, the nominal voltage trajectory converges to the respective equilibrium for any $\tilde{z}(0) \in \tilde{\mc{Z}} \coloneqq \tilde{\mc{Z}}_1 \times \tilde{\mc{Z}}_2 \times \dots \times \tilde{\mc{Z}}_n$. Furthermore, Prop. \ref{prop:PI_set} guarantees that $\mc{S}$ is a robust positive invariant set, such that for all $e(0) \in \mc{S}$ and any $\delta P \in \mbb{W}_{P}$, $\delta Q \in \mbb{W}_{Q}$, the respective solution remains within the set $\mc{S}$ for all $t>0$. Therefore, the above allows us to adopt a driving-driven cascaded system approach, see \cite{sepulchre2012constructive}, where the cascaded dynamics are an interconnection between a driving asymptotically stable subsystem and a driven bounded subsystem. Combining the above results, $\Omega$ is attractive for the solution of the cascaded dynamics $x(t)=\left(e(t),\tilde{z}(t),\sigma_d(t),\sigma_q(t)\right)$ and the solution converges to  $\Omega$ in finite time. Hence, we conclude that the true voltage network system is asymptotically stable with $\lim_{t \to \infty} v(t) = z^o + \left(\hat{z},0\right)^{\top} + e(t) $.
\end{proof}
Furthermore, the following Corollary can be deduced by leveraging the fact that there exist a single unique equilibrium set within the interior of $\mbb{V}$.
\begin{corollary}
An estimate of the RoA of (\ref{eq:Omega_set}) is the set $\ts{int}(\mbb{V})$.
\end{corollary}
The algorithm describing the methodology of tuning the proposed controller is outlined bellow.

\begin{algorithm}[H]
  \caption{Control Design Procedure}
  \begin{algorithmic}[1]
    \STATE Specify desired quadratic error bound $\bar{e} \in \mbb{R}_{>0}$.
    \STATE Specify $\tilde{z}_{\ts{m},i}$, $\delta_i$, and $\gamma_i$ according to Cor. \ref{cor:nom_bound} and \ref{cor:voltage_bound}.
    \STATE Design $K_{d,i}$ according to (\ref{eq:K_d_condition}) and $M_i = \tilde{z}_{\ts{m},i} K_{d,i}$.
    \STATE Design feedback gain $K_i$ according to (\ref{eq:K_bound}).
  \end{algorithmic}
\end{algorithm}

\section{Illustrative Example}\label{sec:Sims}

\begin{figure}[!t]
	\centering
	\resizebox{1\columnwidth}{!}{%
		%
\begin{circuitikz}[american]
%
%
\draw[line width=1mm] (-1,5) -- (0,5); 
\draw[line width=1mm] (2,5) -- (3,5); 
\draw[line width=1mm] (5,5) -- (6,5); 

\draw[line width=1mm] (5,0) -- (6,0); 
\draw[line width=1mm] (2,0) -- (3,0); 
\draw[line width=1mm] (-1,0) -- (0,0); 

\draw[line width=1mm] (6.7,2) -- (6.7,3); 
\draw[line width=1mm] (-1.7,2) -- (-1.7,3); 

\node at (-1.1,5.1){1}; 
\node at (1.9,5.1){2}; 
\node at (4.9,5.1){3}; 
\node at (4.9,-0.1){4}; 
\node at (1.9,-0.1){5}; 
\node at (-1.1,-0.1){6}; 
\node at (6.9,3){7}; 
\node at (-1.9,3){8}; 

\draw (-1.4,6)  node[draw,minimum width=1cm,minimum height=0.5cm] (load1) {CPL}; 
\draw (1.8,6)  node[draw,minimum width=1cm,minimum height=0.5cm] (load2) {CPL}; 
\draw (5,6)  node[draw,minimum width=1cm,minimum height=0.5cm] (load3) {CPL}; 

\draw (-1.4,-1)  node[draw,minimum width=1cm,minimum height=0.5cm] (load4) {CPL}; 
\draw (1.8,-1)  node[draw,minimum width=1cm,minimum height=0.5cm] (load5) {CPL}; 
\draw (5,-1)  node[draw,minimum width=1cm,minimum height=0.5cm] (load6) {CPL}; 

\draw (8,2.3)  node[draw,minimum width=1cm,minimum height=0.5cm] (load7) {CPL}; 
\draw (8,1.7)  node[draw,minimum width=1cm,minimum height=0.5cm] (load71) {\ts{ren}}; 

\draw (-3,2.3)  node[draw,minimum width=1cm,minimum height=0.5cm] (load8) {CPL}; 
\draw (-3,1.7)   node[draw,minimum width=1cm,minimum height=0.5cm] (load81) {\ts{ren}}; 

\draw[thick] (-0.8,5) --++(0,1) --++ (-0.1,0); 
\draw[thick] (2.1,5) --++ (0,0.75);
\draw[thick] (5.2,5) --++ (0,0.75);

\draw[thick] (5.2,0) -- (5.2,-0.75);
\draw[thick] (2.1,0) -- (2.1,-0.75); 
\draw[thick] (-0.8,0) -- (-0.8,-1) --++ (-0.1,0);

\draw[thick] (-1.7,2.4) --++ (-0.8,0);
\draw[thick] (6.7,2.4) --++ ( 0.8,0);


\draw[thick] (-0.4,5) -- (-0.4,6);
\draw[thick] (-0.4,0) -- (-0.4,-1);

\draw[thick] (2.8,0) -- (2.8,-1);
\draw[thick] (2.8,5) -- (2.8,6);

\draw[thick] (5.8,0) --++ (0,-1);
\draw[thick] (5.8,5) --++ (0,1);

\draw[thick] (-1.7,2.8) --++ (-0.8,0);
\draw[thick] (-1.7,2.1) --++ (-0.4,0) --++ (0,-0.3) --++ (-0.4,0);

\draw[thick] (6.7,2.8) --++ (0.8,0);
\draw[thick] (6.7,2.1) --++ (0.4,0) --++ (0,-0.3) --++ (0.4,0);

\draw[thick] (5.8,5) --++ (0,1);

\draw (-0.2,6.5)  node[draw,minimum width=1cm,minimum height=1cm] (inv1) {}; 
\draw (-0.1,6.5) node[npn,scale=0.5 ](npn){};

\draw (2.9,6.5)  node[draw,minimum width=1cm,minimum height=1cm] (inv2) {}; 
\draw (3,6.5) node[npn,scale=0.5 ](npn){};

\draw (6.1,6.5)  node[draw,minimum width=1cm,minimum height=1cm] (inv3) {}; 
\draw (6.2,6.5) node[npn,scale=0.5 ](npn){};

\draw (-0.2,-1.5)  node[draw,minimum width=1cm,minimum height=1cm] (inv1) {}; 
\draw (-0.1,-1.5) node[npn,scale=0.5 ](npn){};

\draw (2.9,-1.5)  node[draw,minimum width=1cm,minimum height=1cm] (inv2) {}; 
\draw (3,-1.5) node[npn,scale=0.5 ](npn){};

\draw (6.1,-1.5)  node[draw,minimum width=1cm,minimum height=1cm] (inv3) {}; 
\draw (6.2,-1.5) node[npn,scale=0.5 ](npn){};

\draw (8,3.2)  node[draw,minimum width=1cm,minimum height=1cm] (inv2) {}; 
\draw (8.1,3.2) node[npn,scale=0.5 ](npn){};

\draw (-3,3.2)  node[draw,minimum width=1cm,minimum height=1cm] (inv3) {}; 
\draw (-2.9,3.2) node[npn,scale=0.5 ](npn){};

%

\draw (-0.5,5) --++ (0,-2) to [cute inductor,l_=$L_{16}$] ++(0,-1) to[/tikz/circuitikz/bipoles/length=7mm,R,l_=$r_{16}$] ++ (0,-1) --++(0,-1);
\draw (2.3,5) --++ (0,-2) to [cute inductor,l_=$L_{25}$] ++(0,-1) to[/tikz/circuitikz/bipoles/length=7mm,R,l_=$r_{25}$] ++ (0,-1) --++(0,-1);
\draw (5.6,5) --++ (0,-2) to [cute inductor,l=$L_{34}$] ++(0,-1) to[/tikz/circuitikz/bipoles/length=7mm,R=$r_{34}$] ++ (0,-1) --++(0,-1);

\draw (-0.3,5) --++ (0,-0.8) to [cute inductor,l^=$L_{12}$]++ (1.2,0) to[/tikz/circuitikz/bipoles/length=7mm,R,l^=$r_{12}$] ++ (1.2,0) --++(0,0.8);
\draw (2.8,5) --++ (0,-0.8) to [cute inductor,l^=$L_{23}$]++ (1.16,0) to[/tikz/circuitikz/bipoles/length=7mm,R,l^=$r_{23}$] ++ (1.2,0) --++(0,0.8);
\draw (2.8,0) --++ (0,0.8) to [cute inductor,l_=$L_{54}$]++ (1,0) to[/tikz/circuitikz/bipoles/length=7mm,R,l_=$r_{54}$] ++ (1.3,0) --++(0,-0.8);

\draw (-0.2,0) --++ (0,3.2) to [cute inductor,l^=$L_{62}$]++ (1,0) to[/tikz/circuitikz/bipoles/length=7mm,R,l^=$r_{62}$] ++ (1.4,0) --++(0,1.8);
\draw (2.6,5) --++ (0,-1.8) --++(0.33,0) to [cute inductor,l^=$L_{24}$]++ (1,0) to[/tikz/circuitikz/bipoles/length=7mm,R,l^=$r_{24}$] ++ (1,0) --++(0.4,0) --++(0,-3.2);
\draw (2.6,0) --++ (0,1.8) --++(0.33,0) to [cute inductor,l^=$L_{35}$]++ (1,0) to[/tikz/circuitikz/bipoles/length=7mm,R,l^=$r_{35}$] ++ (1,0) --++(0.2,0) to[crossing] ++(0.3,0) --++(0,3.2);

\draw (-1.7,2.6) --++ (0.2,0) --++(0,0.33) to [cute inductor,l^=$L_{18}$]++ (0,1) to[/tikz/circuitikz/bipoles/length=7mm,R,l^=$r_{18}$] ++ (0,1) --++(0.6,0) --++(0,0.1);
\draw (-1.7,2.4) --++ (0.2,0) --++(0,-0.33) to [cute inductor,l_=$L_{68}$]++ (0,-1) to[/tikz/circuitikz/bipoles/length=7mm,R,l_=$r_{68}$] ++ (0,-1) --++(0.6,0) --++(0,-0.1);

\draw (6.7,2.6) --++ (-0.2,0) --++(0,0.33) to [cute inductor,l_=$L_{37}$]++ (0,1) to[/tikz/circuitikz/bipoles/length=7mm,R,l_=$r_{37}$] ++ (0,1) --++(-0.6,0) --++(0,0.1);
\draw (6.7,2.4) --++ (-0.2,0) --++(0,-0.33) to [cute inductor,l^=$L_{68}$]++ (0,-1) to[/tikz/circuitikz/bipoles/length=7mm,R,l^=$r_{68}$] ++ (0,-1) --++(-0.6,0) --++(0,-0.1);


\draw (0.3,6.75)--++(0.2,0) to [/tikz/circuitikz/bipoles/length=0.5cm,C,l=$v_{\ts{dc},1}$] ++(0,-0.5) --++(-0.2,0);

\draw (3.4,6.75)--++(0.2,0) to [/tikz/circuitikz/bipoles/length=0.5cm,C,l=$v_{\ts{dc},2}$] ++(0,-0.5) --++(-0.2,0);

\draw (6.6,6.75)--++(0.2,0) to [/tikz/circuitikz/bipoles/length=0.5cm,C,l=$v_{\ts{dc},3}$] ++(0,-0.5) --++(-0.2,0);

\draw (0.3,-1.25)--++(0.2,0) to [/tikz/circuitikz/bipoles/length=0.5cm,C,l=$v_{\ts{dc},6}$] ++(0,-0.5) --++(-0.2,0);

\draw (3.4,-1.25)--++(0.2,0) to [/tikz/circuitikz/bipoles/length=0.5cm,C,l=$v_{\ts{dc},5}$] ++(0,-0.5) --++(-0.2,0);

\draw (6.6,-1.25)--++(0.2,0) to [/tikz/circuitikz/bipoles/length=0.5cm,C,l=$v_{\ts{dc},4}$] ++(0,-0.5) --++(-0.2,0);

\draw (8.5,3.5)--++(0.2,0) to [/tikz/circuitikz/bipoles/length=0.5cm,C,l=$v_{\ts{dc},7}$] ++(0,-0.5) --++(-0.2,0);
\draw (-3.5,3.5)--++(-0.2,0) to [/tikz/circuitikz/bipoles/length=0.5cm,C,l_=$v_{\ts{dc},8}$] ++(0,-0.5) --++(0.2,0);


%
%
%
%
%
%

\end{circuitikz}
        }
        \caption{Single-phase equivalent of the adopted meshed Microgrid network with eight local buses.} 
        \label{fig:SimNetwork} 
\end{figure}
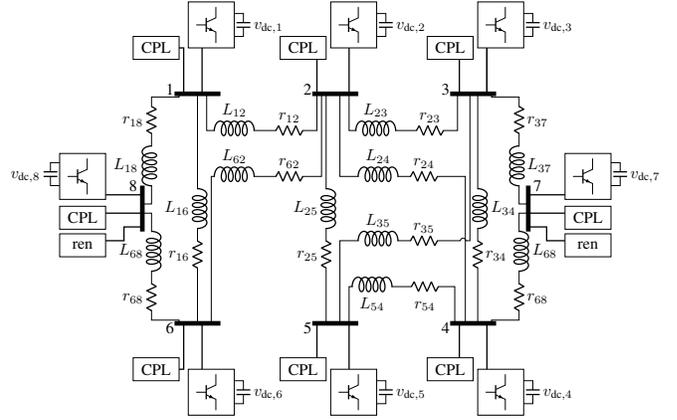
This section will demonstrate the operation of the closed-loop system. An 8-bus AC Microgrid network is considered with topology depicted in Fig. \ref{fig:SimNetwork}. The Microgrid consists of local CPLs, while two nodes include separate power injection from renewable sources. The complete list of the network parameters is shown in Table~\ref{tab:Sims}. We choose $\bar{e} = 2 V$ which results in $\left| \norm{v_i} - \norm{z_i}\right| \leq \sqrt{\bar{e}}$, \ie the maximum distance between the true and nominal local RMS voltage is bounded as $\max\left\{\ts{dist}\left(\norm{v_i},\norm{z_i}\right)\right\} \leq \sqrt{\bar{e}} \ V$. Then, the boundary on the feedback gain from Prop. \ref{prop:PI_set} is computed as $\beta_i = 5.53$ and we set $K_i = \beta_i$ for all $i \in \mc{M}$. 
\begin{table}[t]
\caption{Network component and controller parameters.}
\label{tab:Sims}
\centering
\begin{tabular}{c|c}
Parameter & Value \\
\hline
C [$\mu F$] & $\{ 200, \ 150,\ 180,$\\ 
& $\ 150,\ 150,\ 160 \}$\\
\hline
$R_{ij}$[$\Omega$],$\ (i,j) \in \mc{E}$ & $\{1.5,\ 0.7,\ 1,\ 0.9,\ 1.2,$\ \\
& $ 1.1,\ 1.4,\ 2,\ 1.6  \}$\\
\hline
$L_{ij}$[$H$],$\ (i,j) \in \mc{E}$ & $\{0.2,\ 0.21,\ 0.2,\ 0.3,\ 0.25,$\\
&$ 0.22,\ 0.2,\ 0.22,\ 0.23  \}$\\
\hline
$\omega_g$[$rad/s$] & $2\pi 50$ \\
\hline
$\left(\bar{P}\ts{[W]},\ \bar{Q} \ts{[Var]}, \delta P\ts{[W]}, \ \delta Q \ts{[Var]} \right) $ & $ \left( 600 , \ 500 ,\ 500 , \ 400 \right)$ \\
\hline
Tube width & $2 \sqrt{\bar{e}} = 2.83 V$ \\
\hline
$\mc{Z}_i$ & $\{92 \ts{ V} \leq \norm{z_i} \leq 128 \ts{ V} \}$\\
\end{tabular}
\end{table}
\subsection{Two Node Interaction}
\begin{figure}[!t]
	\centering
		\includegraphics[trim={0.5cm 8cm 2cm 8.5cm},clip,width=0.8\columnwidth]{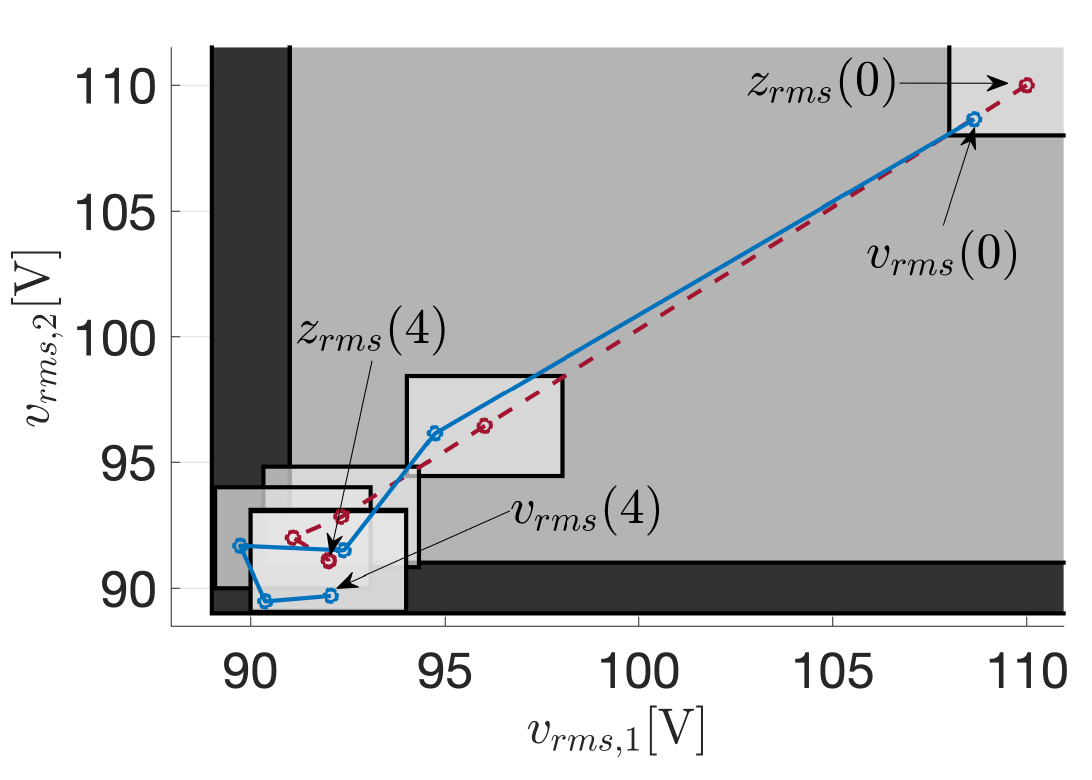}
        \caption{True/nominal voltage trajectories for Node 1 and 2. Cross-sections of the tube are depicted in different time instants. $\mc{Z}_1 \times \mc{Z}_2$ is depicted by light gray, while the true voltage constraint set $\mbb{V}_1 \times \mbb{V}_2$ by dark grey.}
        \label{fig:voltage_state_space} 
\end{figure}
%
Initially, we consider a two-node scenario where the nominal voltage is regulated from $z_{\ts{rms},1}= z_{\ts{rms},2} = 110\ts{ V}$ via  subsequent reference changes to values on the boundary of the nodal nominal constraint set. The system trajectory is depicted in Fig. \ref{fig:voltage_state_space}, showing constraint satisfaction at all times for both the true and the nominal voltage subsystems. The tube property of the cascaded system is shown via the tube cross-sections at different time intervals, where the true voltage is always contained in the desired set.

\subsection{Microgrid Network - HIL Validation}
\begin{figure}[!t]
	\centering
		\includegraphics[trim={0cm 2.2cm 0cm 2.2cm},clip,width=0.5\columnwidth]{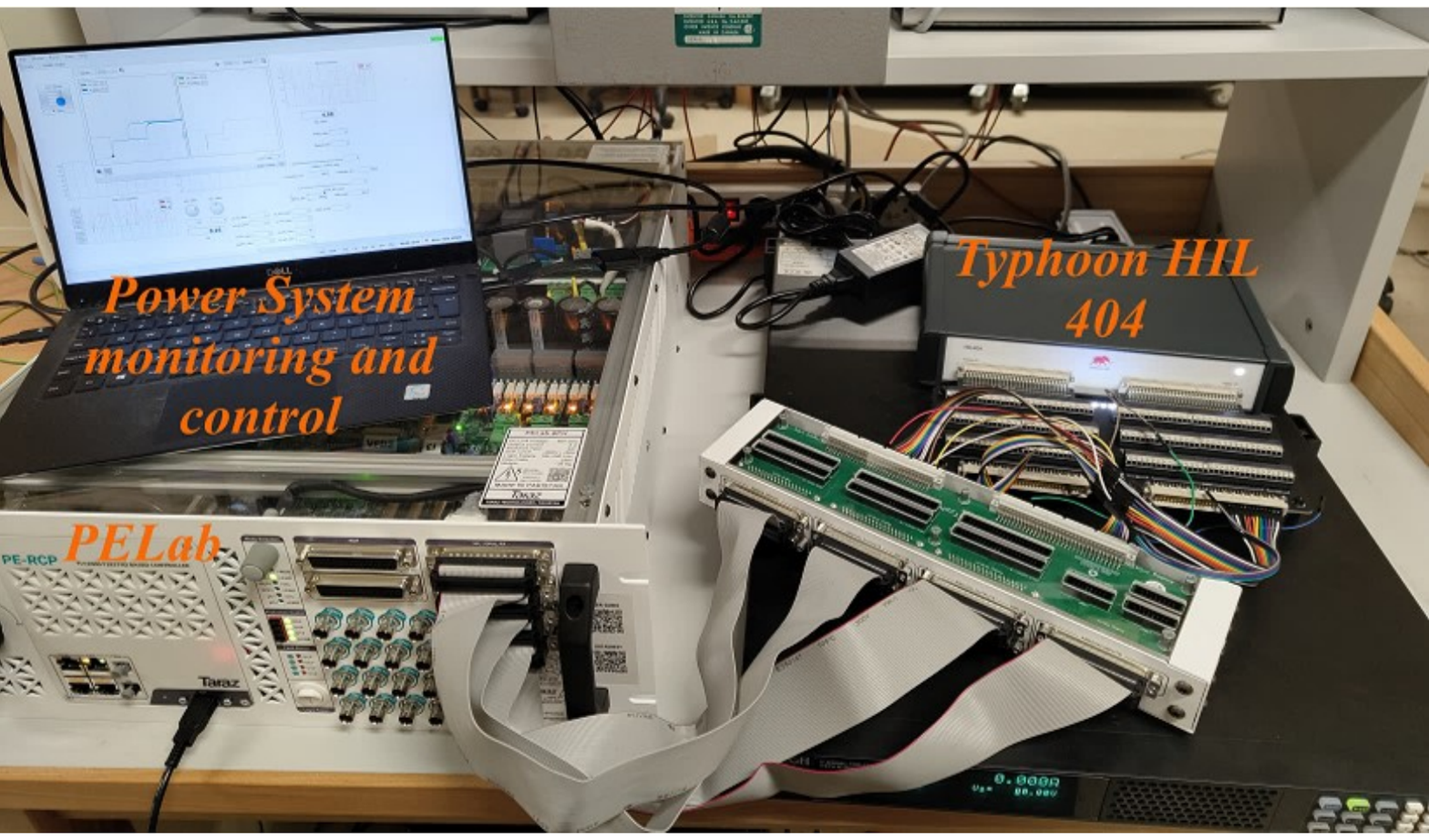}
     %
        \caption{HIL setup.}
        \label{fig:HIL} 
\end{figure}
\begin{figure}[!t]
	\centering
		\includegraphics[trim={0.6cm 9.0cm 2.2cm 10.0cm},clip,width=0.9\columnwidth]{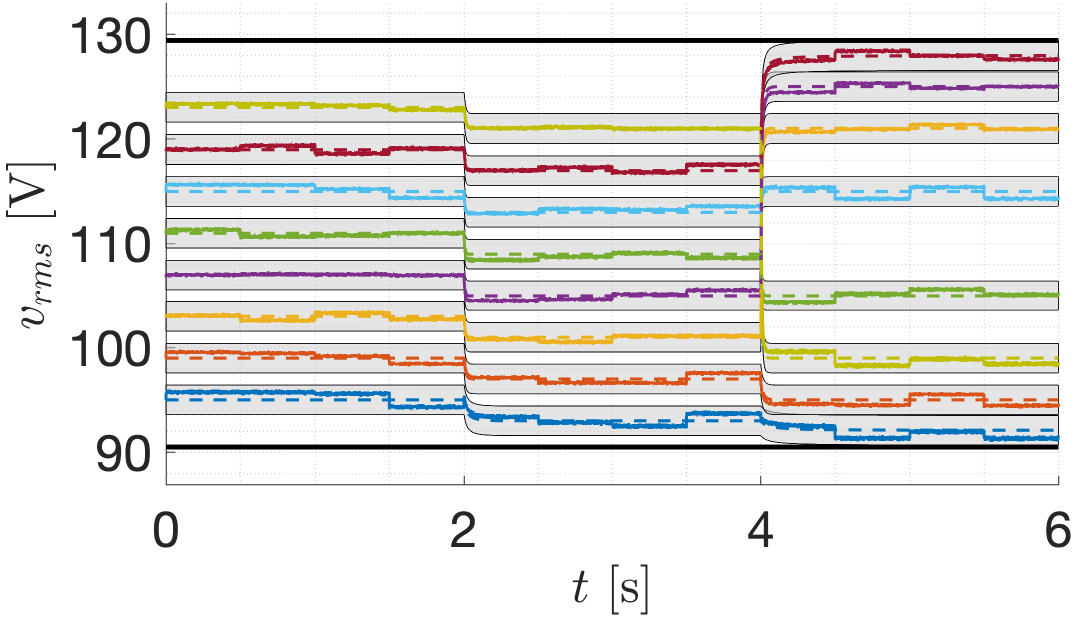}
        \caption{Network true and nominal (dashed) voltage trajectories, constraint set (black solid) and respective tubes (light grey).}
        \label{fig:voltage} 
\end{figure}
\begin{figure}[!t]
	\centering
		\includegraphics[trim={0.6cm 9.3cm 2.2cm 10.2cm},clip,width=0.9\columnwidth]{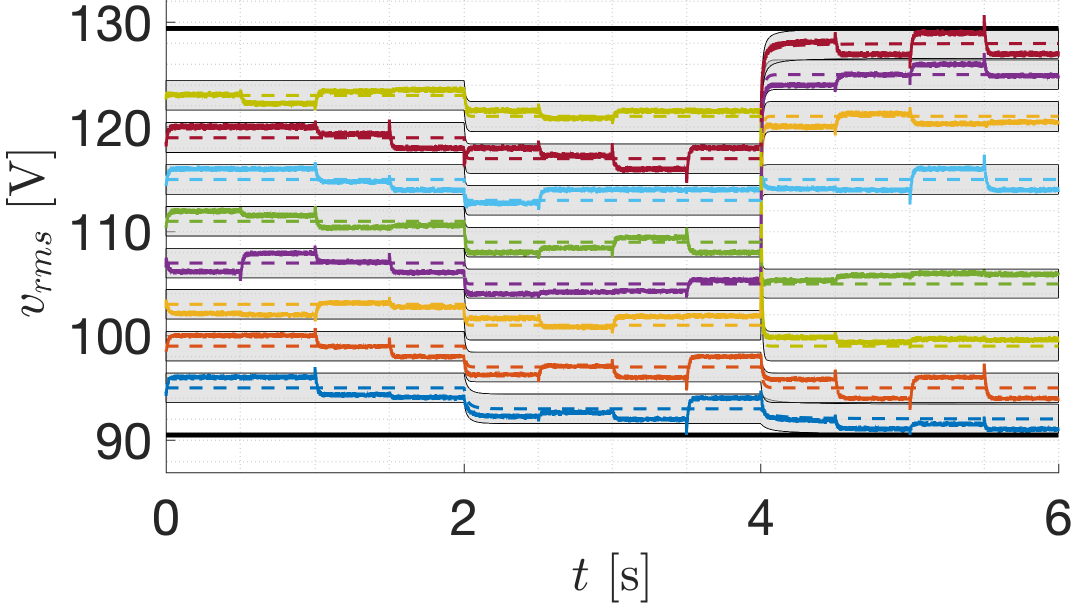}
        \caption{Voltage regulation by conventional droop control.}
        \label{fig:voltage_PI} 
\end{figure}
\begin{figure}[!t]
	\centering
		\includegraphics[trim={0.8cm 10cm 1.2cm 10.5cm},clip,width=0.9\columnwidth]{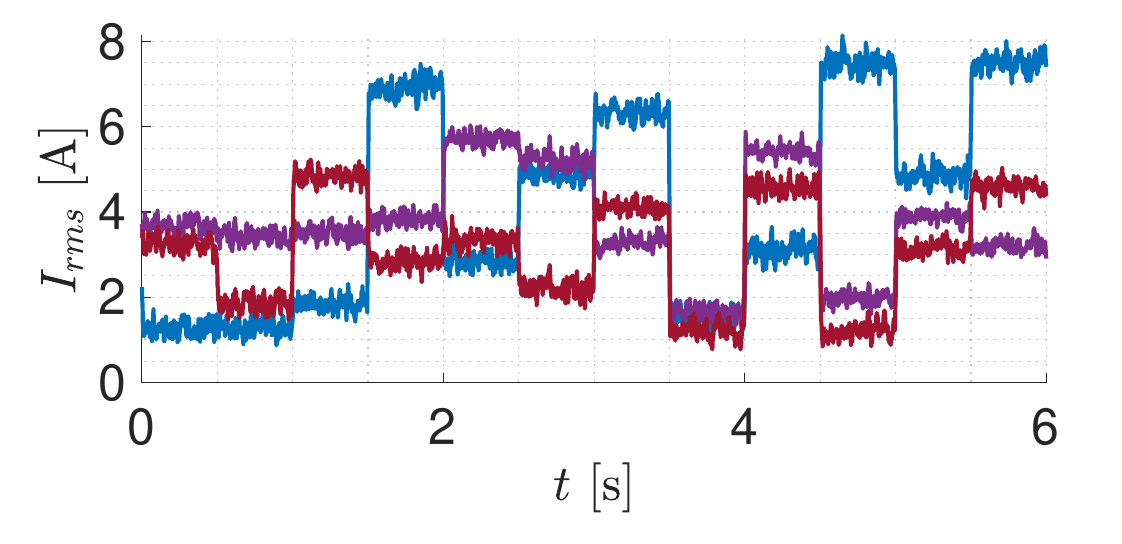}
       \caption{Injected RMS currents for Nodes $(1, 4, 7)$.}
        \label{fig:Current} 
\end{figure}
\begin{figure}[!t]
	\centering
		\includegraphics[trim={0.5cm 10cm 0.5cm 10.8cm},clip,width=0.9\columnwidth]{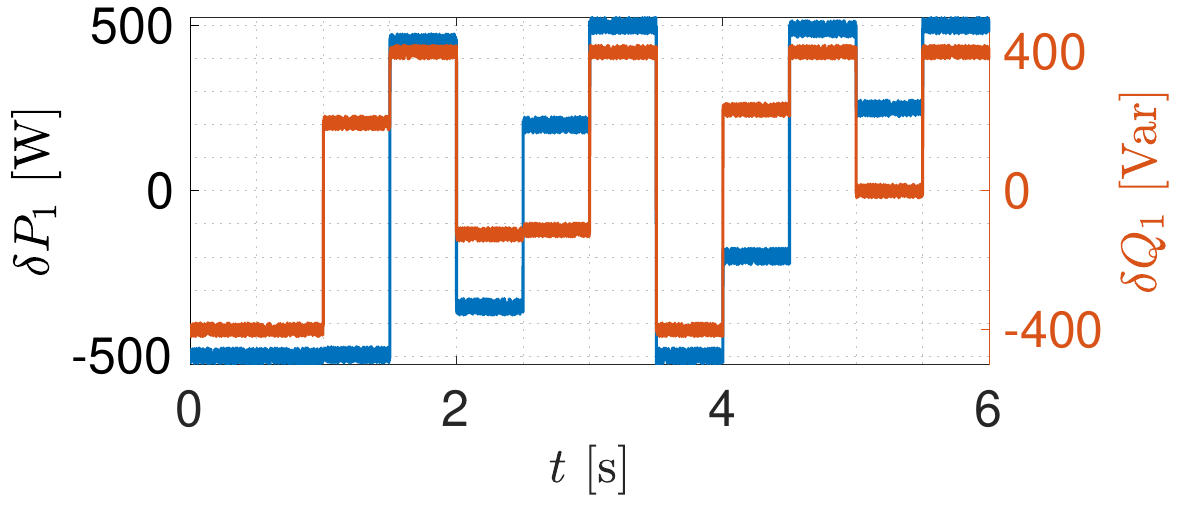}
        \caption{CPL Demand for Node 1.}
        \label{fig:Load} 
\end{figure}
\begin{figure}[!t]
	\centering
		\includegraphics[trim={1cm 10cm 1cm 11cm},clip,width=0.9\columnwidth]{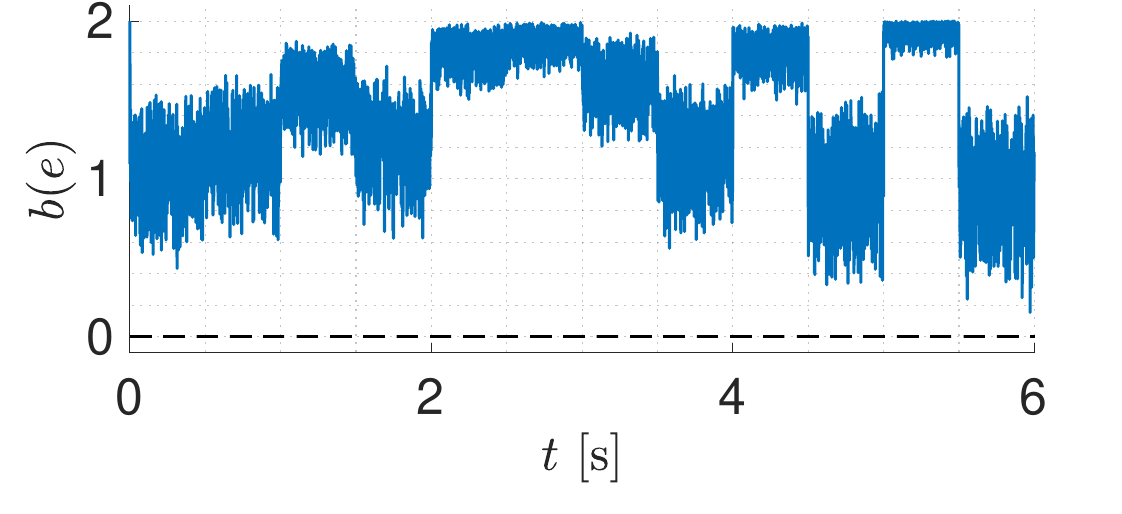}
        \caption{CBF values for Node 1.}
        \label{fig:CBF} 
\end{figure}
Next, we verify the theoretic results of the previous sections via real-time HIL results obtained using a \textit{Typhoon HIL 404} device and an interfacing \textit{PELab} unit from Taraz Technologies, see Fig. \ref{fig:HIL}. Initially, at time $t=0s$, the network system is operating at the equilibrium set, with nominal RMS reference vector $\hat{z}_i = \left\{95\ 99 \ 103 \ 107\ 111\ 115\ 119\ 123\right\} V$, see Fig. \ref{fig:voltage}. Similarly to the previous case, the controller enforces boundedness on the distance between the true and the nominal subsystem trajectories in the respective control tubes and for all load demand perturbations. Then, at time $t=2s$, a nominal reference change occurs with $\hat{z}_i = \left\{93\ 97\ 101\ 105 \ 109\ 113\ 117\ 121\right\}V$, and the network converges to the new equilibrium set. A second reference change happens at $t=4s$, where, this time, the provided setpoint to Nodes 1 and 2 lie outside the constraint set, with $\hat{z}_i = \left\{92\ 95\ 121\ 125\ 105\ 115 \ 129 \ 99 \right\} V$. The nominal controller enforces boundedness in $\mc{Z}_i$, while the true voltage remains within the respective tube and the set $\mbb{V}$. It is noted that the nodal voltages remain within the respective tubes, despite large CPL fluctuations, reaching $1 pu$ step changes. A subset of the nodal currents are also depicted in Fig. \ref{fig:Current}. In order to quantify the effectiveness of the proposed approach, we compare our approach with the conventional $P-V$ droop controller, where the droop gains are tuned accordingly to achieve $\sqrt{\bar{e}}\ V$ deviation from the nominal value at maximum load demand. The results are illustrated in Fig. \ref{fig:voltage_PI}, where the network voltages violate the constraint sets. Considering the load perturbations of Node 1 in Fig. \ref{fig:Load}, the evolution of the control barrier function over the trajectory is depicted in Fig. \ref{fig:CBF}. It is observed that the function remains positive despite the presence of maximum disturbance magnitudes. This also reveals the degree of conservativeness on the bound of feedback gain, stemming from deriving a worst-case scenario condition; the value of the CBF does not reach zero on maximum CPL fluctuations but obtains strictly positive values, albeit close to zero. Ultimately, the proposed controller successfully bounds the true voltage within the desired constraint region, demonstrating system robustness to both inconsistent setpoints and load demand perturbations.

\section{Conclusions}\label{sec:Conclusions}

This study investigates the voltage regulation problem of an AC Microgrid network with high penetration of CPLs. Necessary and sufficient conditions are analytically derived to guarantee boundedness of the network voltages within a tube centred at a nominal trajectory, where the tube is defined by suitable control barrier functions. The mathematical analysis of this paper provided an insight on the interaction between the nonlinear load and the proposed controller parameters. This methodology allowed a characterization of the CPL effect on the closed-loop system vector field. Subsequently, this knowledge was exploited to derive appropriate conditions on the magnitude of the controller tuning parameters such that the network system admits asymptotically stable equilibria and satisfies the desired operational constraints. Future works aim to address the conservativeness on the size of the feedback gain, stemming from deriving a constant, time-invariant, bounding function and assuming a worst-case scenario for the nominal voltage and load demand values. Furthermore, the development of an adaptive tube width according to the load demand will be investigated, in order to accommodate actuation limitations in the form of bounds on the injected currents and address extreme load conditions such that Assumption \ref{ass:load_demand} is violated.

\bibliography{IEEEabrv,library}

\begin{thebibliography}{10}
\providecommand{\url}[1]{#1}
\csname url@samestyle\endcsname
\providecommand{\newblock}{\relax}
\providecommand{\bibinfo}[2]{#2}
\providecommand{\BIBentrySTDinterwordspacing}{\spaceskip=0pt\relax}
\providecommand{\BIBentryALTinterwordstretchfactor}{4}
\providecommand{\BIBentryALTinterwordspacing}{\spaceskip=\fontdimen2\font plus
\BIBentryALTinterwordstretchfactor\fontdimen3\font minus
  \fontdimen4\font\relax}
\providecommand{\BIBforeignlanguage}[2]{{%
\expandafter\ifx\csname l@#1\endcsname\relax
\typeout{** WARNING: IEEEtran.bst: No hyphenation pattern has been}%
\typeout{** loaded for the language `#1'. Using the pattern for}%
\typeout{** the default language instead.}%
\else
\language=\csname l@#1\endcsname
\fi
#2}}
\providecommand{\BIBdecl}{\relax}
\BIBdecl

\bibitem{dorfler2023control}
F.~D{\"o}rfler and D.~Gro{\ss}, ``Control of low-inertia power systems,''
  \emph{Annual Review of Control, Robotics, and Autonomous Systems}, vol.~6,
  no.~1, pp. 415--445, 2023.

\bibitem{chandorkar2002control}
M.~C. Chandorkar, D.~M. Divan, and R.~Adapa, ``Control of parallel connected
  inverters in standalone ac supply systems,'' \emph{IEEE transactions on
  industry applications}, vol.~29, no.~1, pp. 136--143, 2002.

\bibitem{d2015virtual}
S.~D’Arco, J.~A. Suul, and O.~B. Fosso, ``A virtual synchronous machine
  implementation for distributed control of power converters in smartgrids,''
  \emph{Electric Power Systems Research}, vol. 122, pp. 180--197, 2015.

\bibitem{gross2019effect}
D.~Gro{\ss}, M.~Colombino, J.-S. Brouillon, and F.~D{\"o}rfler, ``The effect of
  transmission-line dynamics on grid-forming dispatchable virtual oscillator
  control,'' \emph{IEEE Transactions on Control of Network Systems}, vol.~6,
  no.~3, pp. 1148--1160, 2019.

\bibitem{alfaro2021distributed}
C.~Alfaro, R.~Guzman, L.~G. De~Vicuna, H.~Komurcugil, and H.~Martin,
  ``Distributed direct power sliding-mode control for islanded ac microgrids,''
  \emph{IEEE transactions on industrial electronics}, vol.~69, no.~10, pp.
  9700--9710, 2021.

\bibitem{khan2022robust}
A.~Khan, M.~M. Khan, Y.~Li, J.~Chuanwen, M.~U. Shahid, and I.~Khan, ``A robust
  control scheme for voltage and reactive power regulation in islanded ac
  microgrids,'' \emph{Electric Power Systems Research}, vol. 210, p. 108179,
  2022.

\bibitem{mohiuddin2019droop}
S.~M. Mohiuddin and J.~Qi, ``Droop-free distributed control for ac microgrids
  with precisely regulated voltage variance and admissible voltage profile
  guarantees,'' \emph{IEEE Transactions on Smart Grid}, vol.~11, no.~3, pp.
  1956--1967, 2019.

\bibitem{markovic2021understanding}
U.~Markovic, O.~Stanojev, P.~Aristidou, E.~Vrettos, D.~Callaway, and G.~Hug,
  ``Understanding small-signal stability of low-inertia systems,'' \emph{IEEE
  Transactions on Power Systems}, vol.~36, no.~5, pp. 3997--4017, 2021.

\bibitem{he2023nonlinear}
X.~He, V.~H{\"a}berle, I.~Suboti{\'c}, and F.~D{\"o}rfler, ``Nonlinear
  stability of complex droop control in converter-based power systems,''
  \emph{IEEE Control Systems Letters}, vol.~7, pp. 1327--1332, 2023.

\bibitem{subotic2020lyapunov}
I.~Suboti{\'c}, D.~Gro{\ss}, M.~Colombino, and F.~D{\"o}rfler, ``A lyapunov
  framework for nested dynamical systems on multiple time scales with
  application to converter-based power systems,'' \emph{IEEE Transactions on
  Automatic Control}, vol.~66, no.~12, pp. 5909--5924, 2020.

\bibitem{michos2023robust}
G.~Michos, P.~R. Baldivieso-Monasterios, G.~C. Konstantopoulos, and P.~A.
  Trodden, ``Robust two-layer control of dc microgrids with fluctuating
  constant power load demand,'' \emph{IEEE Transactions on Control of Network
  Systems}, vol.~11, no.~1, pp. 427--438, 2023.

\bibitem{michos2024dynamic}
G.~Michos, G.~C. Konstantopoulos, and P.~A. Trodden, ``Dynamic tube control for
  dc microgrids,'' \emph{IEEE Control Systems Letters}, 2024.

\bibitem{braitor2024voltage}
A.-C. Braitor, H.~Siguerdidjane, and A.~Iovine, ``On the voltage stability and
  network scalability of onboard dc microgrids for hybrid electric aircraft,''
  \emph{IEEE Transactions on Aerospace and Electronic Systems}, 2024.

\bibitem{areerak2017adaptive}
K.~Areerak, T.~Sopapirm, S.~Bozhko, C.~I. Hill, A.~Suyapan, and K.~Areerak,
  ``Adaptive stabilization of uncontrolled rectifier based ac--dc power systems
  feeding constant power loads,'' \emph{IEEE Transactions on Power
  Electronics}, vol.~33, no.~10, pp. 8927--8935, 2017.

\bibitem{al2017constant}
M.~K. AL-Nussairi, R.~Bayindir, S.~Padmanaban, L.~Mihet-Popa, and P.~Siano,
  ``Constant power loads (cpl) with microgrids: Problem definition, stability
  analysis and compensation techniques,'' \emph{Energies}, vol.~10, no.~10, p.
  1656, 2017.

\bibitem{molinas2008constant}
M.~Molinas, D.~Moltoni, G.~Fascendini, J.~A. Suul, and T.~Undeland, ``Constant
  power loads in ac distribution systems: An investigation of stability,'' in
  \emph{2008 IEEE international symposium on industrial electronics}.\hskip 1em
  plus 0.5em minus 0.4em\relax IEEE, 2008, pp. 1531--1536.

\bibitem{liu2012novel}
Z.~Liu, J.~Liu, W.~Bao, Y.~Zhao, and F.~Liu, ``A novel stability criterion of
  ac power system with constant power load,'' in \emph{2012 Twenty-Seventh
  Annual IEEE Applied Power Electronics Conference and Exposition
  (APEC)}.\hskip 1em plus 0.5em minus 0.4em\relax IEEE, 2012, pp. 1946--1950.

\bibitem{molinas2008investigation}
M.~Molinas, D.~Moltoni, G.~Fascendini, J.~Suul, R.~Faranda, and T.~Undeland,
  ``Investigation on the role of power electronic controlled constant power
  loads for voltage support in distributed ac systems,'' in \emph{2008 IEEE
  Power Electronics Specialists Conference}.\hskip 1em plus 0.5em minus
  0.4em\relax IEEE, 2008, pp. 3597--3602.

\bibitem{liu2014infinity}
Z.~Liu, J.~Liu, W.~Bao, and Y.~Zhao, ``Infinity-norm of impedance-based
  stability criterion for three-phase ac distributed power systems with
  constant power loads,'' \emph{IEEE Transactions on Power Electronics},
  vol.~30, no.~6, pp. 3030--3043, 2014.

\bibitem{mansoorhoseini2022islanded}
P.~Mansoorhoseini, B.~Mozafari, and S.~Mohammadi, ``Islanded ac/dc microgrids
  supervisory control: A novel stochastic optimization approach,''
  \emph{Electric Power Systems Research}, vol. 209, p. 108028, 2022.

\bibitem{zhou2021distributed}
J.~Zhou, H.~Sun, Y.~Xu, R.~Han, Z.~Yi, L.~Wang, and J.~M. Guerrero,
  ``Distributed power sharing control for islanded single-/three-phase
  microgrids with admissible voltage and energy storage constraints,''
  \emph{IEEE Transactions on Smart Grid}, vol.~12, no.~4, pp. 2760--2775, 2021.

\bibitem{KOLSCH202012229}
\BIBentryALTinterwordspacing
L.~Kölsch, K.~Wieninger, S.~Krebs, and S.~Hohmann, ``Distributed frequency and
  voltage control for ac microgrids based on primal-dual gradient
  dynamics⁎⁎this work was funded by the deutsche forschungsgemeinschaft
  (dfg, german research foundation)—project number 360464149.''
  \emph{IFAC-PapersOnLine}, vol.~53, no.~2, pp. 12\,229--12\,236, 2020, 21st
  IFAC World Congress. [Online]. Available:
  \url{https://www.sciencedirect.com/science/article/pii/S2405896320314877}
\BIBentrySTDinterwordspacing

\bibitem{mottaghizadeh2022robust}
M.~Mottaghizadeh, F.~Aminifar, T.~Amraee, and M.~Sanaye-Pasand, ``Robust fuzzy
  model predictive control for voltage regulation in islanded microgrids,''
  \emph{IET Generation, Transmission \& Distribution}, vol.~16, no.~5, pp.
  1013--1029, 2022.

\bibitem{11037272}
T.~E. Kavvathas and G.~C. Konstantopoulos, ``Novel virtual synchronous
  generator design with dynamic virtual inertia and bounded frequency, current
  and voltage characteristics,'' \emph{IEEE Transactions on Sustainable
  Energy}, pp. 1--14, 2025.

\bibitem{rocabert2012control}
J.~Rocabert, A.~Luna, F.~Blaabjerg, and P.~Rodriguez, ``Control of power
  converters in ac microgrids,'' \emph{IEEE transactions on power electronics},
  vol.~27, no.~11, pp. 4734--4749, 2012.

\bibitem{qoria2018tuning}
T.~Qoria, F.~Gruson, F.~Colas, X.~Guillaud, M.-S. Debry, and T.~Prevost,
  ``Tuning of cascaded controllers for robust grid-forming voltage source
  converter,'' in \emph{2018 Power Systems Computation Conference
  (PSCC)}.\hskip 1em plus 0.5em minus 0.4em\relax IEEE, 2018, pp. 1--7.

\bibitem{xin2016synchronous}
H.~Xin, L.~Huang, L.~Zhang, Z.~Wang, and J.~Hu, ``Synchronous instability
  mechanism of pf droop-controlled voltage source converter caused by current
  saturation,'' \emph{IEEE Transactions on Power Systems}, vol.~31, no.~6, pp.
  5206--5207, 2016.

\bibitem{ames2016control}
A.~D. Ames, X.~Xu, J.~W. Grizzle, and P.~Tabuada, ``Control barrier function
  based quadratic programs for safety critical systems,'' \emph{IEEE
  Transactions on Automatic Control}, vol.~62, no.~8, pp. 3861--3876, 2016.

\bibitem{xu2015robustness}
X.~Xu, P.~Tabuada, J.~W. Grizzle, and A.~D. Ames, ``Robustness of control
  barrier functions for safety critical control,'' \emph{IFAC-PapersOnLine},
  vol.~48, no.~27, pp. 54--61, 2015.

\bibitem{wu2024fixed}
Z.~Wu and H.~Cheng, ``Fixed-time control of distributed secondary voltage and
  frequency for microgrids considering state-constrained,'' \emph{Electrical
  Engineering}, vol. 106, no.~3, pp. 2637--2649, 2024.

\bibitem{hassan2024resilience}
K.~Hassan, D.~Selvaratnam, and H.~Sandberg, ``On resilience guarantees by
  finite-time robust control barrier functions with application to power
  inverter networks,'' \emph{IEEE Open Journal of Control Systems}, 2024.

\bibitem{kundu2019distributed}
S.~Kundu, S.~Geng, S.~P. Nandanoori, I.~A. Hiskens, and K.~Kalsi, ``Distributed
  barrier certificates for safe operation of inverter-based microgrids,'' in
  \emph{2019 American Control Conference (ACC)}.\hskip 1em plus 0.5em minus
  0.4em\relax IEEE, 2019, pp. 1042--1047.

\bibitem{habibi2021unfalsified}
S.~I. Habibi and A.~Bidram, ``Unfalsified switching adaptive voltage control
  for islanded microgrids,'' \emph{IEEE Transactions on Power Systems},
  vol.~37, no.~5, pp. 3394--3407, 2021.

\bibitem{ullah2022voltage}
S.~Ullah, L.~Khan, I.~Sami, and J.~Ro, ``Voltage/frequency regulation with
  optimal load dispatch in microgrids using smc based distributed cooperative
  control,'' \emph{IEEE Access}, vol.~10, pp. 64\,873--64\,889, 2022.

\bibitem{redheffer1972theorems}
R.~Redheffer, ``The theorems of bony and brezis on flow-invariant sets,''
  \emph{The American Mathematical Monthly}, vol.~79, no.~7, pp. 740--747, 1972.

\bibitem{espinoza1992voltage}
J.~Espinoza, G.~Joos, and P.~Ziogas, ``Voltage controlled current source
  inverters,'' in \emph{Proceedings of the 1992 International Conference on
  Industrial Electronics, Control, Instrumentation, and Automation}.\hskip 1em
  plus 0.5em minus 0.4em\relax IEEE, 1992, pp. 512--517.

\bibitem{Khalil2002a}
H.~K. Khalil and J.~W. Grizzle, \emph{Nonlinear systems}.\hskip 1em plus 0.5em
  minus 0.4em\relax Prentice hall Upper Saddle River, NJ, 2002, vol.~3.

\bibitem{baldivieso2020constrained}
P.~R. Baldivieso-Monasterios and G.~C. Konstantopoulos, ``Constrained control
  for microgrids with constant power loads,'' in \emph{2020 59th IEEE
  Conference on Decision and Control (CDC)}.\hskip 1em plus 0.5em minus
  0.4em\relax IEEE, 2020, pp. 3341--3346.

\bibitem{agundis2018extended}
G.~Agundis-Tinajero, N.~L.~D. Aldana, A.~C. Luna, J.~Segundo-Ramirez,
  N.~Visairo-Cruz, J.~M. Guerrero, and J.~C. Vazquez,
  ``Extended-optimal-power-flow-based hierarchical control for islanded ac
  microgrids,'' \emph{IEEE Transactions on Power Electronics}, vol.~34, no.~1,
  pp. 840--848, 2018.

\bibitem{arkhangelski2021day}
J.~Arkhangelski, M.~Abdou-Tankari, and G.~Lefebvre, ``Day-ahead optimal power
  flow for efficient energy management of urban microgrid,'' \emph{IEEE
  transactions on industry applications}, vol.~57, no.~2, pp. 1285--1293, 2021.

\bibitem{michos2023control}
G.~Michos, G.~C. Konstantopoulos, P.~A. Trodden, and V.~Kadirkamanathan,
  ``Control of isolated ac microgrids with constant power loads: A set
  invariance approach,'' in \emph{2023 31st Mediterranean Conference on Control
  and Automation (MED)}.\hskip 1em plus 0.5em minus 0.4em\relax IEEE, 2023, pp.
  239--244.

\bibitem{axler2024linear}
S.~Axler, \emph{Linear algebra done right}.\hskip 1em plus 0.5em minus
  0.4em\relax Springer Nature, 2024.

\bibitem{tisseur2001quadratic}
F.~Tisseur and K.~Meerbergen, ``The quadratic eigenvalue problem,'' \emph{SIAM
  review}, vol.~43, no.~2, pp. 235--286, 2001.

\bibitem{sepulchre2012constructive}
R.~Sepulchre, M.~Jankovic, and P.~V. Kokotovic, \emph{Constructive nonlinear
  control}.\hskip 1em plus 0.5em minus 0.4em\relax Springer Science \& Business
  Media, 2012.

\end{thebibliography}

\end{document}